%% file: main.tex
\documentclass[sigconf, balance=false]{acmart}
\usepackage{subcaption}
\usepackage{popets}
\usepackage[ruled,vlined]{algorithm2e}
\usepackage{multirow}
\usepackage{placeins}

\setcopyright{popets}
\copyrightyear{2026}

\acmYear{2026}
\acmVolume{2026}
\acmNumber{4}
\acmDOI{10.56553/popets-2026-0131}
\acmISBN{}
\acmConference{Proceedings on Privacy Enhancing Technologies}
\input{macro}

\begin{document}
\sloppy
\title{Poison to Detect: Detection of Targeted Overfitting in Federated Learning}

%% \citi{} and \state{} are fully optional.
%% You do not need to provide values for the \city{}, \state{}, and \country{} commands 
%% within the \author{} commands, but YOU MUST leave the blank commands in the file.
\author{Soumia Zohra El Mestari}
\authornote{Both authors contributed equally to this research.}
\orcid{0000-0002-1399-605X}
\affiliation{%
  \institution{University of Luxembourg}
  \city{Luxembourg}
  \country{Luxembourg}
}
\email{soumia.elmestari@uni.lu}

\author{Maciej Krzysztof Zuziak}
\authornotemark[1]
\orcid{0000-0003-4297-4973}
\affiliation{%
  \institution{University of Leeds}
  \city{Leeds}
  \country{United Kingdom}}
\email{M.Zuziak@leeds.ac.uk}

\author{Gabriele Lenzini}
\orcid{0000-0001-8229-3270}
\affiliation{%
  \institution{University of Luxembourg}
  \city{Luxembourg}
  \country{Luxembourg}
}
\email{gabriele.lenzini@uni.lu}

%\renewcommand{\shortauthors}{El Mestari et al.}
\renewcommand{\shortauthors}{El Mestari, Zuziak, and Lenzini}

\begin{abstract}
 Federated Learning (FL) enables collaborative model training among clients without centralising data, making it a widely adopted privacy-enhancing technology (PET). Despite its privacy benefits, FL remains vulnerable to orchestrator-driven privacy attacks. In this paper, we study an underexplored threat in which a dishonest orchestrator intentionally manipulates the aggregation process to induce targeted overfitting in local models of specific clients. Although prior work focuses on reducing information leakage during training, we emphasise early client-side detection of targeted overfitting, allowing clients to disengage before significant harm occurs. To this end, we propose three detection techniques—label flipping, backdoor trigger injection, and model fingerprinting—which enable clients to verify the integrity of the global aggregation. We evaluated our methods across multiple datasets and attack scenarios. In single-client attacks, all three methods detect orchestrator-induced overfitting within 1–2 training rounds with F1 scores up to  0.7. Scalability experiments further show that detection effectiveness is influenced by cohort composition and method parameters. These results demonstrate that client-side integrity testing can provide early, effective, and scalable detection, supporting safer deployment of FL systems.
\end{abstract}


\maketitle

\section{Introduction}
Federated Learning (FL) is a distributed training paradigm that enables multiple clients to collaboratively train a model under the coordination of a central orchestrator without the need to share their data among them or with the orchestrator \cite{kairouz2021advances,li2020federated}. FL is often considered a privacy-preserving alternative to centralised training. This setting enables the global model to leverage various data sources, thus overcoming the limitations of processing sensitive data that cannot be shared between local environments. However, despite these advantages, decentralisation introduces a new attack vector, where both clients and the orchestrator can become malicious actors targeting one another. Although previous research has focused mainly on attacks mounted by malicious clients that manipulate local and global updates and even local data \cite{tolpegin2020data}, less attention has been paid to orchestrator-driven attacks. The aggregation process in FL, though ostensibly neutral, can be exploited to induce specific learning behaviours in targeted clients, a strategy that we refer to as \textit{targeted aggregation}. In this context, the orchestrator can manipulate model updates to strategically amplify or suppress specific data characteristics, which may inadvertently degrade model performance or, even worse, cause a model to memorise or leak sensitive client data, threatening the integrity and privacy of the system \cite{bagdasaryan2020backdoor}. 
Among the possible instantiations of targeted aggregation, our work focuses on one we term \textit{targeted overfitting}, which may also extend to the strategic selection of clients to amplify or suppress particular data patterns. In this scenario, the orchestrator performs a double aggregation process where one aggregation is benign, including all updates received from clients, and its output is sent to the non-targeted clients, and a second aggregation is malicious, where the orchestrator aggregates using updates received from a subset of targeted clients, and the result of this malicious aggregation is shared with the targeted clients only. %Over iterations, this malicious aggregation leads the targeted clients to overfit on their data. By overfitting the targeted client’s model to its local data, the orchestrator increases the risk of Membership Inference Attacks (MIA) \cite{Shokri2017MIA}, where the adversary can infer the presence of specific data points within the training set of a particular targeted client. In addition to that, overfitted models may memorise specific data points \cite{lin2024overmemorization}, increasing their vulnerability to Data Reconstruction Attacks \cite{balle2022reconstructing,bertran2024reconstruction,panchendrarajan2021dataset} and Model Inversion Attacks \cite{Fredrikson2015ACM_SIGSAC, He2019ACM, Wu2016AMF}. 

Over iterations, this causes targeted clients to overfit their local data, increasing their susceptibility to Membership Inference Attacks (MIA)\cite{Shokri2017MIA}, Data Reconstruction Attacks\cite{balle2022reconstructing,bertran2024reconstruction,panchendrarajan2021dataset}, and Model Inversion Attacks\cite{Fredrikson2015ACM_SIGSAC, He2019ACM, Wu2016AMF}.

%Over multiple iterations, the targeted client’s model becomes more tailored to its local data distribution, reducing its performance on unseen data. This causes the target local models to diverge from the global one. 
Over multiple iterations, the targeted client's model diverges from the global one, losing generalisation ability. 
Furthermore, Backdoor Attacks (BA) can be exacerbated by overfitting the target client's model to specific trigger patterns %, thereby embedding malicious behaviours more effectively 
\cite{bagdasaryan2020backdoor}. Additionally, based on the BA strategies of Fang \etal \cite{Fang2020Backdoorn}, targeted overfitting may offer a pathway for training hijacking, wherein the orchestrator manipulates model updates to align with adversarial objectives, effectively steering the targeted client's model toward specific, potentially harmful patterns. 
% Novelty and contribution  
\begin{figure}
    \centering
    \includegraphics[width=1\linewidth]{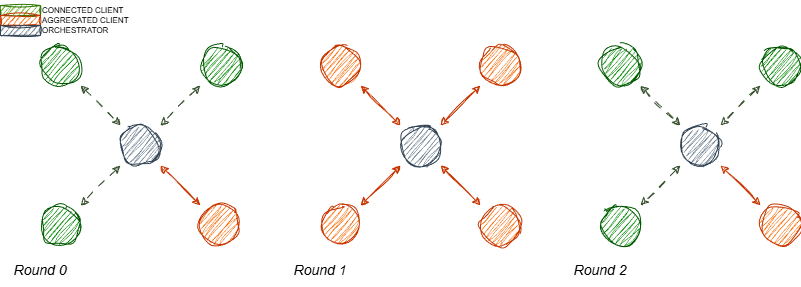}
    
    \caption{Targeted overfitting demonstration. The target client (bottom-left) repeatedly engages in both honest and malicious aggregation, allowing the adversary to mimic honest FL while obtaining a vulnerable local model. Non-sampled clients are coloured green, aggregated clients are orange.}
    \label{fig:targ_overfit_vis}

\end{figure}

\textbf{Contribution} Unlike previous work that focuses on preventing targeted aggregation, our study addresses the detection aspect of the problem. Defence mechanisms like Differential Privacy (DP) \cite{Muah2021ICASSP} impose a utility-privacy tradeoff that persists even in the absence of any attack; a detection mechanism, by contrast, avoids this cost entirely or informs when deploying such prevention mechanisms is warranted. To this end, we propose three client-side detection frameworks that enable clients in an FL setting to autonomously identify targeted overfitting induced by the orchestrator: label flipping detection, backdoor trigger detection, and gradient/weight fingerprinting. Operating without inter-client communication or cooperation, our methods target timely detection of orchestrator-driven attacks that pose significant privacy risks, including membership inference and data reconstruction.

\begin{figure*}[ht]
    \centering
    \includegraphics[width=0.7\textwidth,height=7cm]{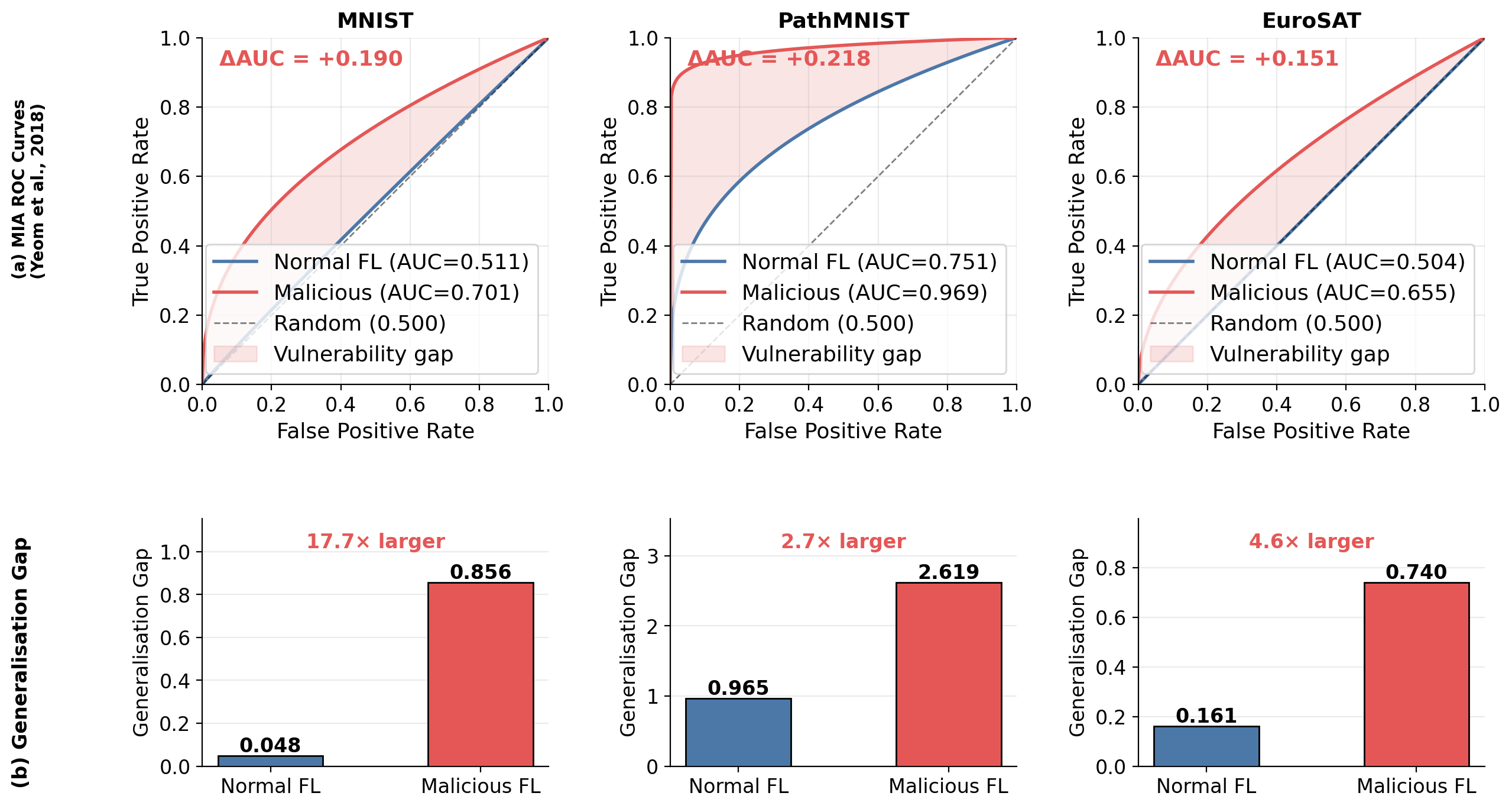}

        \caption{Targeted overfitting amplifies membership 
    inference vulnerability. Row (a) shows MIA ROC curves under 
    normal FL and malicious aggregation across three datasets. 
    Row (b) shows the corresponding generalisation gap. }
    \label{fig:mia_summary}
\end{figure*}

\section{Related works}
\paragraph{\textbf{Orchestrator as an attacker}} Attacks altering model integrity in FL have focused mainly on client-to-client or client-to-orchestrator threats \cite{Blanchard_NIPS2017,Fang_2020_localmodelpoison,bhagoji2019analyzing, gupta2023novel,Zhang_2022_FLDETECTOR}.
 In client-driven attacks, adversarial participants may manipulate local updates to corrupt the global model (e.g., model poisoning attacks \cite{Blanchard_NIPS2017} or insert backdoors \cite{bagdasaryan2020backdoor}). Some works have explored collusion among clients to extract information about other clients or to degrade the performance of the global model \cite{Nasr2019IEESP}. 
However, the case of an adversarial orchestrator remains comparatively under-explored despite its privileged access to local models and the aggregation process. Wang \etal \cite{wang2019beyond} demonstrated that a malicious server can exploit federated updates using a GAN-based discriminator to reconstruct data from specific targeted clients, highlighting the severity of server-side attacks. Lyu \etal \cite{Lyu2020} further emphasised that a malicious or even “honest-but-curious” server poses a significant threat to the integrity of the FL system.
Such a server can send custom models to targeted clients~\cite{Chen_2022_FL_survey}, manipulate training dynamics~\cite{ Lyu2020}, and bias aggregation by favouring selected updates~\cite{Zhang_2024_survey}.

Some server-side attacks can also be part of a multi-step attack pipeline aimed at compromising client privacy. Such attacks — including Membership Inference Attacks (MIA) \cite{Yeom_2018_CSF, Shokri2017MIA}, Attribute Inference Attacks (AIA) \cite{Yeom_2018_CSF}, Model Inversion Attacks \cite{yang2025MI}, can all be exacerbated when target models overfit their training data, a vulnerability that a malicious orchestrator can subtly induce. By manipulating global updates, the orchestrator can increase the generalisation gap, thereby enhancing the effectiveness of inference attacks.\vspace{0.5em}

\textbf{Overfitting} 
Client's model may overfit its local data, increasing its susceptibility to privacy attacks such as Source Inference Attacks (SIA) from a curious server \cite{zhao2025finp}. Additionally, data heterogeneity—where client data are non-IID —can lead to local model overfitting to their specific distribution, causing divergence from the global objective and hindering convergence \cite{yang2022robust, rothchild2020fetchsgd}. Overfitting is also exploited in client-side poisoning attacks, where a malicious client intentionally overfits on poisoned samples to introduce backdoors into the global model \cite{zawad2021curse}. Moreover, rigid training configurations, such as applying a fixed number of local epochs to all clients regardless of dataset size or complexity, can result in uneven training—causing some clients to be underfitted and others to be overfitted—ultimately degrading the quality of the aggregated global model \cite{huang2020loadaboost}.

\textbf{Detection methods}
Client-side detection of orchestrator-driven attacks remains largely under-explored, though most existing techniques are not explicitly designed to detect a malicious server; several principles from existing research can be adapted. For example, local performance monitoring can serve as an early warning system: by maintaining a private test set, a client can compare training and testing performance locally. Persistent overfitting after aggregation may suggest orchestrator-driven manipulation, though it can also arise naturally from data heterogeneity~\cite{zhang2023survey}. Another promising approach is collaborative validation, as proposed in defences against backdoor attacks \cite{andreina2021baffle}, where randomly selected clients validate the integrity of the global model using their private data and collectively vote on potential anomalies. This mechanism could be generalised to detect other server-side manipulations, particularly when multiple clients report similar degradation or instability patterns. Furthermore, strategies aimed at reducing local overfitting — such as promoting fairness-in-privacy through collaborative regularisation \cite{zhao2025finp} — represent a complementary direction that could be adapted for detecting orchestrator-induced overfitting. However, in this work, we employ a threat model that is more challenging from the detection perspective, assuming that clients do not cooperate with each other, making collaborative methods such as those presented in \cite{andreina2021baffle} infeasible to implement.

\section{Problem statement}
\subsection{Threat model}
\subsubsection{Adversary’s capabilities} We consider a threat model applicable to horizontal FL. 
We assume that the orchestrator is modelled as a malicious adversary that can tamper with the orchestration and aggregation protocols. We therefore assume that the orchestrator can maliciously modify the initial aggregation protocol in order to selectively sample a targeted client or clients, causing them to overfit the model to local datasets. Moreover, we assume no external cross-communication between clients. Hence, the orchestrator is the only entity that can exchange messages with local nodes.

\subsubsection{Adversary’s objective} The adversary's objective is to cause clients to over-fit their local models by sampling targeted clients repeatedly. The adversary's main goal is to ultimately receive models that are memorising data points stored locally, in order to amplify its attack capabilities %and ability 
to infer information about sensitive data.

\subsubsection{Targeted implementations} The adversary targets clients by tampering with the orchestration protocol. We implement FedAvg~\cite{mcmahan2017communication} and FedOpt~\cite{reddi2021adaptive} as the main aggregation protocols. The threat model generalises to most federated aggregation methods, though specific protocols may affect the ease of inducing targeted overfitting.

\subsubsection{Clients modelling} Clients adopt a believing-but-cautious stance: they participate in standard federated training while seeking to verify whether aggregation is honest. We assume no inter-client cooperation. Detection is the primary objective; we do not prescribe mitigation strategies, leaving clients to decide how to act on the information obtained.
\subsection{Targeted overfitting}

As described above, the malicious aggregator (\ie the orchestration server in a horizontal FL setting) intentionally performs modified aggregations in order to make a subset of clients overfit in their local datasets. The targeted clients can be either a single client or a subset of clients among the elected clients. 

Suppose that the local models obtained from the clients selected in iteration $t$ are denoted by the set $\{ W_i^t \}_{i=1}^{N}$, where $N$ is the number of clients participating in that round. Let $\mathcal{S}^t = \{1, 2, \ldots, N\}$ represent the set of selected clients in iteration $t$. In a benign FL setting, the orchestrator aggregates all local updates received from $\mathcal{S}^t$ using an aggregation function $\mathcal{A}(\cdot)$, such as FedAvg \cite{mcmahan2017communication}. The resulting global model at iteration $t$ is then given by:
\[
W_{\text{global}}^t = \mathcal{A}\left( \{ W_i^t \}_{i \in \mathcal{S}^t} \right).
\]
However, to model the orchestrator’s behaviour in our study, we consider a malicious setting in which the orchestrator may deviate from standard protocol by aggregating updates from only a subset of the selected clients. Let $\mathcal{S}_\text{mal}^t \subset \mathcal{S}^t$ denote a subset of clients of size $n < N$, selectively chosen by the malicious server. The corresponding malicious aggregation is defined as:
\[
\tilde{W}_{\text{global}}^t = \mathcal{A}\left( \{ W_i^t \}_{i \in \mathcal{S}_\text{mal}^t} \right).
\]
This biased aggregation may serve various adversarial objectives, such as model manipulation, targeted overfitting, or privacy leakage, by excluding benign updates or overemphasising the influence of certain clients.

Compared to direct server-side manipulations, targeted overfitting is an especially effective adversarial strategy because it hides within the normal mechanics of FL. Direct attacks—such as arbitrary parameter rewriting, gradient tampering, or model replacement—tend to be structurally detectable: they require elevated server privileges, often violate integrity checks, and produce parameter updates that deviate sharply from expected training behaviour. In contrast, targeted overfitting relies only on selectively excluding benign updates, using the same aggregation pathway that honest training already depends on. The resulting model drift resembles natural variability caused by non-IID client data, making it statistically camouflaged rather than syntactically malicious. Since FL intentionally trusts locally trained updates, the preference of a single client on the server appears indistinguishable from benign distributional heterogeneity. This makes targeted overfitting both harder to detect and more privacy-threatening, as it induces memorisation in precisely the clients the adversary wants to compromise.

In this study, we analyse the targeted overfitting strategies under two distinct scenarios, which arise from two modes of targeting.

The modes correspond to the adversary targeting either a single client or a fixed subset of $n$ clients.
Formally, the scenarios are defined as follows:

\begin{itemize}
    \item \textbf{Scenario I:} Targeting a specific single client $i^* \in \mathcal{S}^t$ at every iteration $t$, i.e.,
    \[
        \mathcal{S}_\text{mal}^t = \{i^*\}, \quad \forall t.
    \]
    \item \textbf{Scenario II:} Targeting a fixed subset $\mathcal{S}_{\text{mal}}$ at every iteration, i.e.,
    \[
        \mathcal{S}_\text{mal}^t = \mathcal{S}_{\text{mal}}, \quad \forall t.
    \]
\end{itemize}

We focus on these two scenarios because they capture the fundamental behaviours of targeted overfitting. Scenario I (Figure \ref{fig:targ_overfit_vis}) represents the continuous targeting of a single client, while Scenario II generalises this idea to targeting a fixed subset of clients. 

Other possible scenarios include targeting a single client or a subset of clients periodically during a fixed number of iterations, which can be viewed as special cases or episodic instances of these two scenarios. Thus, studying Scenarios I and II allows us to analyse the core effect of targeted overfitting, as well as the proposed detection mechanisms in a clear and generalisable way.

\subsection{Privacy consequences of targeted overfitting}
\label{sec:privacy_consequences}

Targeted overfitting introduces direct, measurable privacy risks to the affected clients. As shown in Appendix~\ref{Appendix_Memorisation}, the generalisation gap between the target client’s model and non-targeted clients increases progressively across rounds under malicious aggregation, suggesting an increased memorisation of the target client’s training data. To evaluate whether this behaviour translates into a concrete privacy threat, we apply the threshold-based membership inference attack (MIA) of Yeom et al.~\cite{Yeom_2018_CSF} to the target client’s model under both honest and malicious aggregation across three datasets. Figure~\ref{fig:mia_summary} reports both the MIA AUC and the corresponding generalisation gap. Under honest aggregation, MIA AUC remains close to random for MNIST and EUROSAT ($0.511$ and $0.504$ respectively), indicating limited inference leakage in these settings. In contrast, PATHMNIST exhibits higher baseline vulnerability even without an attack ($0.751$ AUC). Under malicious aggregation, MIA AUC increases consistently across all datasets, reaching $0.701$, $0.655$, and $0.969$ for MNIST, EUROSAT, and PATHMNIST, respectively ($\Delta = +0.190$, $+0.151$, and $+0.218$). These increases are accompanied by substantial growth in the generalisation gap, which rises from $0.048$ to $0.856$ for MNIST, from $0.161$ to $0.740$ for EUROSAT, and from $0.965$ to $2.619$ for PATHMNIST. 
Overall, under malicious aggregation, MIA AUC increases by up to $+0.218$, while the generalisation gap increases by as much as $17.7\times$ relative to honest aggregation, confirming that targeted overfitting measurably amplifies privacy exposure. 
These findings motivate the need for early client-side detection mechanisms before memorisation accumulates over successive training rounds.

\section{Detecting targeted overfitting}

Client-level distance thresholding—comparing received updates to the dispatched global model—is a standard approach for detecting anomalous behaviour. In principle, this can detect targeted overfitting; however, the detection completely fails when the orchestrator uses adaptive optimisers (e.g., Adam, RMSProp), which rely on internal states such as momentum and adaptive learning rates. The internal states cause each global update to shift in direction and magnitude depending on  past gradients, leading to non-zero and variable distances between the dispatched and received models—even in dishonest settings. As a result, distance-based methods become noisy and unreliable. In this section, we propose three methods for detecting targeted overfitting, enabling clients to autonomously identify it at an early stage.
Early detection is crucial because the longer targeted aggregation continues, the more victims overfit on their local data. By identifying suspicious updates promptly, clients can take preventive measures such as opting out of the training process or raising an alert in the system to ensure mitigation occurs before the attack fully manifests.

\subsection{Label flipping poisoning} 

\begin{algorithm}[t] % Note the [H] here to force placement
\SetAlgoLined
\SetKwInOut{Input}{Input}
\SetKwInOut{Output}{Output}
\Input{Local dataset $\mathcal{D}$, flip ratio $\alpha$}
\Output{Clean subset $\mathcal{D}_{\text{clean}}$, flipped subset $\mathcal{D}_{\text{flip}}$}

$\mathcal{F} \gets$ class frequency map over $\mathcal{D}$ \\
$c^* \gets \arg\min_k \mathcal{F}[k]$ \hfill\textcolor{gray}{\# Get the least frequent class} \\

$\mathcal{S} \gets \{(x_i, y_i) \in \mathcal{D} \mid y_i = c^*\}$ \\
$n_{\text{flip}} \gets \lfloor \alpha \cdot |\mathcal{S}| \rfloor$ \hfill\textcolor{gray}{\# Number of samples to flip} \\

Randomly select $\mathcal{S}_{\text{flip}} \subset \mathcal{S}$, $|\mathcal{S}_{\text{flip}}| = n_{\text{flip}}$ %\hfill\textcolor{gray}{\# Subset to flip} 
\\

Define $c' = (c^* + 1) \mod K$ \textcolor{gray}{\# New label %for  
%flipped \\ \hfill samples}
}\\

$\mathcal{D}_{\text{flip}} \gets \{(x, c') \mid (x, y) \in \mathcal{S}_{\text{flip}} \}$ \\
$\mathcal{D}_{\text{clean}} \gets \mathcal{D} \setminus \mathcal{S}_{\text{flip}}$ %\hfill\textcolor{gray}{\# Remaining clean data} 
\\

\textbf{return} $\mathcal{D}_{\text{clean}}, \mathcal{D}_{\text{flip}}$

\caption{Creating Label-Flipping Poisons}
\label{alg:label_flipping}
\end{algorithm}

We propose a controlled label-flipping mechanism that clients can perform to detect targeted overfitting. The core idea is to poison a small subset of the data so that its effect remains locally detectable even after model aggregation. If a client observes that the global model performs almost as well as the local model on the poisoned subset $\mathcal{D}_{\text{flip}}$ (evaluated immediately after local training), it may suggest that the server selectively favoured the client's local model during aggregation (see algorithm \ref{alg:label_flipping}). In other words, under honest aggregation, the global model should "correct" the poisoned signal, resulting in a drop in performance on $\mathcal{D}_{\text{flip}}$.

Specifically, each client selects the least frequent class in its local dataset and flips a fraction $\alpha$ of its samples using a cyclic rule: $\text{flip}(y) = (y + 1) \mod K$.
This cyclic approach ensures that a label \( y \) is always mapped to a different class, avoiding self-flipping. This ensures semantic change while avoiding label collisions. In addition to that, targeting the least frequent class minimises disruption to overall training and reduces the likelihood of triggering server-side defences. The flipped samples form a poisoned subset $\mathcal{D}_{\text{flip}}$, while the rest of the dataset remains clean.  Algorithm \ref{alg:label_flipping} details the steps followed. To quantify this behavior, we define the \textit{Poison Effectiveness Score (PES)}:
\(
\text{PES} = \text{Acc}(M_{\text{local}}, \mathcal{D}_{\text{flip}}) - \text{Acc}(M_{\text{agg}}, \mathcal{D}_{\text{flip}})
\)

where $M_{\text{local}}$ is the client model before aggregation,
$M_{\text{agg}}$ is the aggregated global model, and
$\mathcal{D}_{\text{flip}}$ is the poisoned subset.
Let $\tau_{pes}$ denote an empirically estimated threshold derived from the distribution of PES under honest aggregation.
We interpret $\mathrm{PES} \ge \tau_{pes}$ as consistent with honest aggregation, and $\mathrm{PES} < \tau_{pes}$ as evidence of anomalous retention of the poisoned signal. The threshold $\tau_{pes}$ corresponds to a lower empirical quantile of the reference PES distribution.

\subsection{Backdoor trigger poisoning} 
\begin{algorithm}[h]
\SetAlgoLined
\DontPrintSemicolon
\KwIn{Trigger pattern $\mathcal{T}$, target label $y_{\tau}$,  local dataset $\mathcal{D}_i$, trigger subset $\mathcal{D}_{\text{trigger}} \subset \mathcal{D}_i $, $\theta$ detection threshold }
\KwOut{Decision: Is the client likely to be targeted?}

% Step 1: Create poisoned dataset by injecting the trigger

$\mathcal{D}_i^{\text{poison}} \leftarrow \{ (\text{ApplyTrigger}(\mathbf{x}, \mathcal{T}), y_{\tau}) \;|\; (\mathbf{x}, y) \in \mathcal{D}_{trigger} \}$
%\hfill\textcolor{gray}{\# For each local sample, apply the trigger pattern and assign the target label}

% Step 2: Train local model with poisoned dataset
$w_i^t \leftarrow \text{Train}(\mathcal{D}_i^{clean} \cup \mathcal{D}_i^{\text{poison}})$
%\hfill\textcolor{gray}{\# Train local model with poisoned dataset}

% Step 3: Send local model to orchestrator
Send $w_i^t$ to the orchestrator

% Step 4: Aggregate client models to update global model
$w_{\text{global}}^{t+1} \leftarrow \text{AggregateClientModels}(\{w_i\})$
%\hfill\textcolor{gray}{\# Update global model by aggregating client updates}

% Step 5: Compute trigger influence on global model
%\textcolor{gray}{\# S is the trigger influence on global model} 

$\mathcal{S} \leftarrow \frac{1}{|\mathcal{D}_{\text{trigger}}|} \sum_{\mathbf{x} \in \mathcal{D}_{\text{trigger}}} \mathbb{I}\left( \text{Predict}(w_{\text{global}}^{t+1}, \mathbf{x}) = y_{\tau} \right)$

% Step 6: Compare influence score against threshold
\uIf{$\mathcal{S} \geq \theta$}{
    \textcolor{gray}{\# Detection of selective aggregation based on influence score} \\
    Flag \textbf{selective aggregation} %(i.e., high influence of update)
}
\Else{
    No indication of targeted aggregation
}

\caption{Detection via Backdoor Triggers}
\label{alg:backdoor_trigger}
\end{algorithm}

In this method, clients use a known trigger pattern to poison a small subset of their local data and monitor whether the effect of this trigger persists in the aggregated global model. The intuition is that, under honest aggregation, the influence of a client-specific backdoor should be diluted. If the trigger remains effective after aggregation, it may indicate that the orchestrator disproportionately favoured the poisoned update, suggesting targeted overfitting. Algorithm~\ref{alg:backdoor_trigger} details the detection steps.

\begin{figure}[t]
    \centering
    \includegraphics[width=0.8\columnwidth,height=0.1\textheight,keepaspectratio]{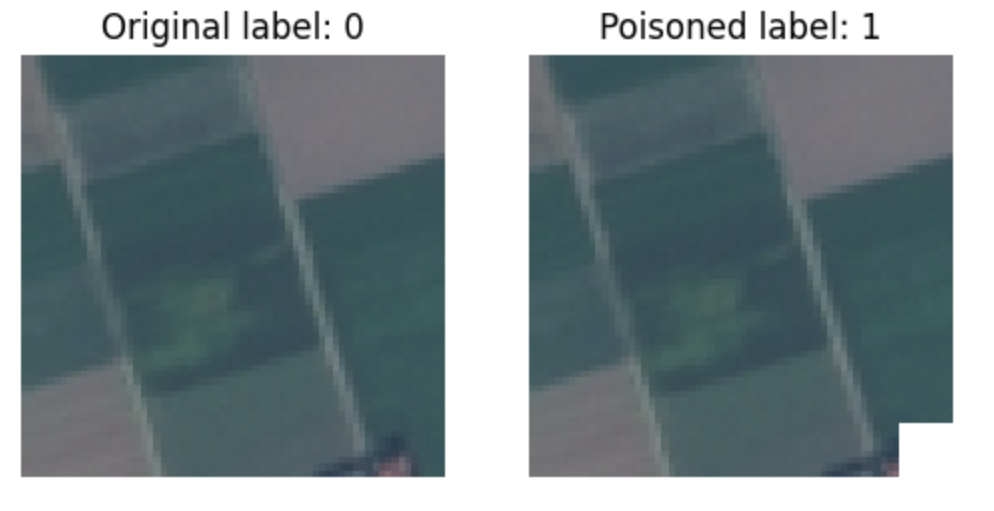}

    \caption{EUROSAT \cite{helber2019eurosat}sample with inserted backdoor trigger (white rectangle in the bottom right corner of the second image).}
    \label{backdoor_example}
\end{figure}

To perform the test, the client selects a predefined trigger pattern \(\mathcal{T}\) and a target label \(y_{\tau}\). Figure \ref{backdoor_example} shows an example of this backdoor on EUROSAT \cite{helber2019eurosat} sample where the $\tau$ is a small white patch at the right bottom corner. A poisoned subset \(\mathcal{D}_i^{\text{poison}}\) is created by applying \(\mathcal{T}\) to some local samples and assigning them the label \(y_{\tau}\). The client trains on the combined dataset \(\mathcal{D}_i \cup \mathcal{D}_i^{\text{poison}}\), then sends the resulting model \(w_i^t\) to the orchestrator.

Once the global model \(w_{\text{global}}^{t+1}\) is received, the client evaluates it on a local trigger evaluation set \(\mathcal{D}_{\text{trigger}}\) and computes the \textit{Trigger Influence Score} \(\mathcal{S}\), defined as the fraction of inputs for which the global model predicts the target label \(y_{\tau}\):

\[
\mathcal{S} = \frac{1}{|\mathcal{D}_{\text{trigger}}|} \sum_{\mathbf{x} \in \mathcal{D}_{\text{trigger}}} \mathbb{I}\left( \text{Predict}(w_{\text{global}}, \mathbf{x}) = y_{\tau} \right)
\]

A low score of \textit{Trigger Influence Score} \(\mathcal{S}\) indicates that the poisoned update has been properly diluted during honest aggregation, whereas a high score suggests that the orchestrator disproportionately favoured the client’s update, allowing the trigger to persist.  
If \(\mathcal{S} \geq \theta\), where \(\theta\) is a predefined threshold, the client flags possible selective aggregation before extensive overfitting occurs.

To select the threshold \(\theta\) we consider two strategies:  

\textbf{(1) Statistical Bound } Without calibration data, we model the number of successful triggers as $\text{Bin}(m,p_0)$,  where $m = |\mathcal{D}_{\text{trigger}}|$ and $p_0$ is the baseline hit-rate (e.g., $1/C$ for $C$ classes). The threshold is  
\[
\theta = \frac{k^*}{m}, \quad 
k^* = \min \{k : \Pr[\text{Bin}(m,p_0)\ge k] \le f_{\text{FP}}\},
\] 
where $f_{\text{FP}}$ is the desired false positive rate and $k$ is the number of successful trigger detections observed in the trigger evaluation set. A normal approximation may be used for large $m$.

\textbf{(2) Empirical Quantile Method.} Given access to benign rounds, we estimate the distribution of benign $\mathcal{S}$ values and set $\theta$ to a high empirical quantile. This quantile acts as a data-driven false-positive control: choosing, for instance, the 95th percentile ensures that, under benign behaviour, fewer than 5\% of rounds exceed $\theta$. This anchors the threshold in observed system variability rather than an arbitrary constant, providing a principled balance between sensitivity to poisoned updates and robustness to natural fluctuations.

\begin{algorithm}[h]
\SetAlgoLined
\DontPrintSemicolon
\KwIn{Local gradient $g$, fingerprint vector $s$, strength scalar of the fingerint vector $\alpha > 0$,  $method \in \{\text{threshold}, \text{history}\}$}
\KwOut{Decision: Is the client likely to be targeted?}

$\tilde{g} \leftarrow g + \alpha s$ \hfill \textcolor{gray}{\# Inject fingerprint into local update} \\

Send $\tilde{g}$ to the orchestrator \;

$G \leftarrow \text{Receive}(G)$ \hfill \textcolor{gray}{\# Receive global update } \\

%$\text{strength} \leftarrow \langle G, s \rangle$ \hfill \textcolor{gray}{\# Compute unnormalised \\
%\hfill fingerprint strength} \\
$\text{strength} \leftarrow \langle G, s \rangle$ \hfill \textcolor{gray}{\# unnormalised fingerprint strength} \\

$\text{similarity} \leftarrow \frac{\langle G, s \rangle}{\|G\| \cdot \|s\|}$ \hfill \textcolor{gray}{\# Optionally normalize via  \\ \hfill cosine similarity} \\

\BlankLine

 \uIf{$m = \text{threshold}$}{
        \uIf{$\text{strength} > \tau_{\text{dot}}$ \textbf{ or } $\text{similarity} > \tau_{\text{cos}}$}{
            Flag \textbf{targeted aggregation} \;
        }
        \Else{
            No indication of targeting \;
        }
    }
    \uElseIf{$m = \text{history}$}{
        \uIf{$\text{strength} \gg \mu$}{
            \textcolor{gray}{\# Compare to historical mean $\mu$} \\
            Flag \textbf{targeted aggregation} \;
        }
        \Else{
            No indication of targeting \;
        }
    }
\caption{Detection via Gradient Fingerprinting }
\label{alg:gradient_fingerprinting}
\end{algorithm}

\subsection{Gradient/weight fingerprinting} 

Our third detection method is the "Gradients/weights fingerprinting". This method relies on fingerprinting the weights or the gradients (algorithm \ref{alg:gradient_fingerprinting}\footnote{The algorithm details the fingerprinting procedure for gradients, we note that we follow the same procedure to inject the fingerprint in the weights, the choice of where to inject the fingerprint depends on the optimisation algorithm used.}), depending on the global optimiser used by the orchestrator. To elaborate, in federated learning, the model updates exchanged between the clients and the orchestrator depend on the global optimiser: algorithms such as FedAvg \cite{mcmahan2017communication} aggregate the model weights after local training, while methods such as FedSGD \cite{mcmahan2017communication} or Scaffold \cite{karimireddy2020scaffold} rely on gradient averaging. The client perturbs its local gradient/weight \( g \) after finishing the local epochs by injecting a small, secret vector \( s \), resulting in a fingerprinted update \( \tilde{g} = g + \alpha s \), where \( \alpha > 0 \) controls the fingerprint strength. The fingerprint vector \( s \) is client-specific and ideally low-correlation with \( g \); common choices include random unit vectors \( (s = v/\|v\|,\, v \sim \mathcal{N}(0, I)) \), sparse binary vectors over client-specific indices, or deterministic vectors derived from the client ID (e.g., \( s = \texttt{HashToVector}(\texttt{ClientID}) \)). The client sends \( \tilde{g} \) to the server, which aggregates updates from multiple clients. Upon receiving the global model update \( G \), the client evaluates the fingerprint strength as \( \text{Strength}_t = \langle G_t, s \rangle \) or optionally its cosine similarity. During typical FL rounds, the fingerprint is expected to be diluted by other clients' updates, resulting in low strength. However, if the client’s update is selectively amplified, the fingerprint will be preserved, leading to an unusually high strength. The value of \( \alpha \) should be small (\(10^{-3} \) to \( 10^{-2}\)) to minimise the influence on training while maintaining detectability. 
\paragraph{\textbf{Detection assessment}}To assess the fingerprint presence post-aggregation, we propose two strategies, namely, threshold-based detection and history-based comparison. 
In the \textbf{\textit{threshold-based method}} (detailed in Algorithm \ref{alg:threshold_fingerprint}), the fingerprint scaling factor $\alpha$, detection thresholds $\tau_{\text{dot}}$, and $\tau_{\text{cos}}$, respectively referring to the dot product threshold and the cosine similarity threshold, are determined analytically before training by considering the model’s dimensionality $d$, fingerprint sparsity $s$, and expected client participation ratio $f$. First, we calculate the total number of model parameters $d$ and the number of non-zero fingerprint elements $k$ determined by the sparsity level $s$. The injection strength 
$\alpha$ is set to achieve a target dot product strength when the fingerprint is fully used. Then, using the expected honest client contribution fraction $f$, the algorithm derives detection thresholds for both dot product $\tau_{\text{dot}}$ and cosine $\tau_{\text{cos}}$ similarity by applying a detection margin multiplier $\gamma$. This approach enables an automatic selection of fingerprint parameters to balance detectability and training influence. However, this approach assumes uniform participation of clients and may be less robust in heterogeneous settings.

In the second strategy (\ie \textbf{\textit{history-based comparison}}), the client compares the current strength to a historical average
\( \mu = \frac{1}{T} \sum_{t=1}^T \langle G_t, s \rangle \). If \( \text{Strength}_t \gg \mu \), this indicates overrepresentation and potential targeting. Hence, the history-based comparison method adapts better to dynamic participation and distribution shifts but requires sufficient historical data and may incur delayed detection. Each approach offers distinct strengths, and combining them can provide more reliable detection across varying federated learning scenarios.

\begin{algorithm}[t]
\caption{Analytical Threshold for Fingerprint}
\label{alg:threshold_fingerprint}
\KwIn{Model $M$, sparsity $s$, target dot strength $S_{\text{target}}$, honest fraction $f$, detection margin $\gamma$}
\KwOut{Fingerprint strength $\alpha$, dot threshold $\tau_{\text{dot}}$, cosine threshold $\tau_{\text{cos}}$}

$d \gets \sum_{p \in M} \text{numel}(p)$\ \textcolor{gray}{\# Number of model parameters} \\

$k \gets \lfloor s \cdot d \rfloor$\  \textcolor{gray}{\# non-zero fingerprint entries} \\

$\alpha \gets \frac{S_{\text{target}}}{k}$\ %\textcolor{gray}{\# fingerprint strength} \\

$S_{\text{honest}} \gets f \cdot S_{\text{target}}$\ \textcolor{gray}{\# estimating honest contribution} \\

$\tau_{\text{dot}} \gets \gamma \cdot S_{\text{honest}}$\;

$\tau_{\text{cos}} \gets \min(1.0, \gamma \cdot f)$\;

\Return $\alpha$, $\tau_{\text{dot}}$, $\tau_{\text{cos}}$

\end{algorithm}

\section{Experiments}
We empirically evaluate our three detection methods across multiple datasets and under the two scenarios described earlier, focusing on detection speed and reliability. We describe the experimental setup, followed by a cross-silo proof-of-concept, scalability experiments, and results overview.

\subsection{Datasets description} \label{dataset_description}
Experiments are conducted on: MNIST \cite{deng2012mnist}, CIFAR10 \cite{krizhevsky2009Cifar10}, CIFAR100 \cite{krizhevsky2009learning}, PATHMNIST \cite{yang2023medmnist} and EUROSAT \cite{helber2019eurosat}. Clients' datasets are generated using standard non-IID Dirichlet partitioning \cite{kairouz2021advances,reddi2021adaptive}. For each client $i \in P$, we first sample a probability vector over classes from a Dirichlet distribution parameterised by $\alpha$, \ie $\boldsymbol{v}_i \sim \mathrm{Dir}(\alpha)$.
This vector $\boldsymbol{v}_i$ specifies the class proportions for client $i$. 
The local dataset $D_i$ is then constructed by drawing samples from the global dataset according to $\boldsymbol{v}_i$. 

\paragraph{\textbf{Cross-silo evaluation}}
For proof-of-concept experiments, we use a small number of clients. In Scenario I, we used five clients, targeting a single client (client 1), while other clients (\eg 0, 2, 3, 4) were honestly aggregated by the orchestrator. In Scenario~II, seven clients are split into four honest and three targeted clients. The slight difference in cardinality is intentional and highlights the sensitivity of detection under targeted overfitting.

 \begin{figure*}
    \centering
    \includegraphics[width=1\linewidth]{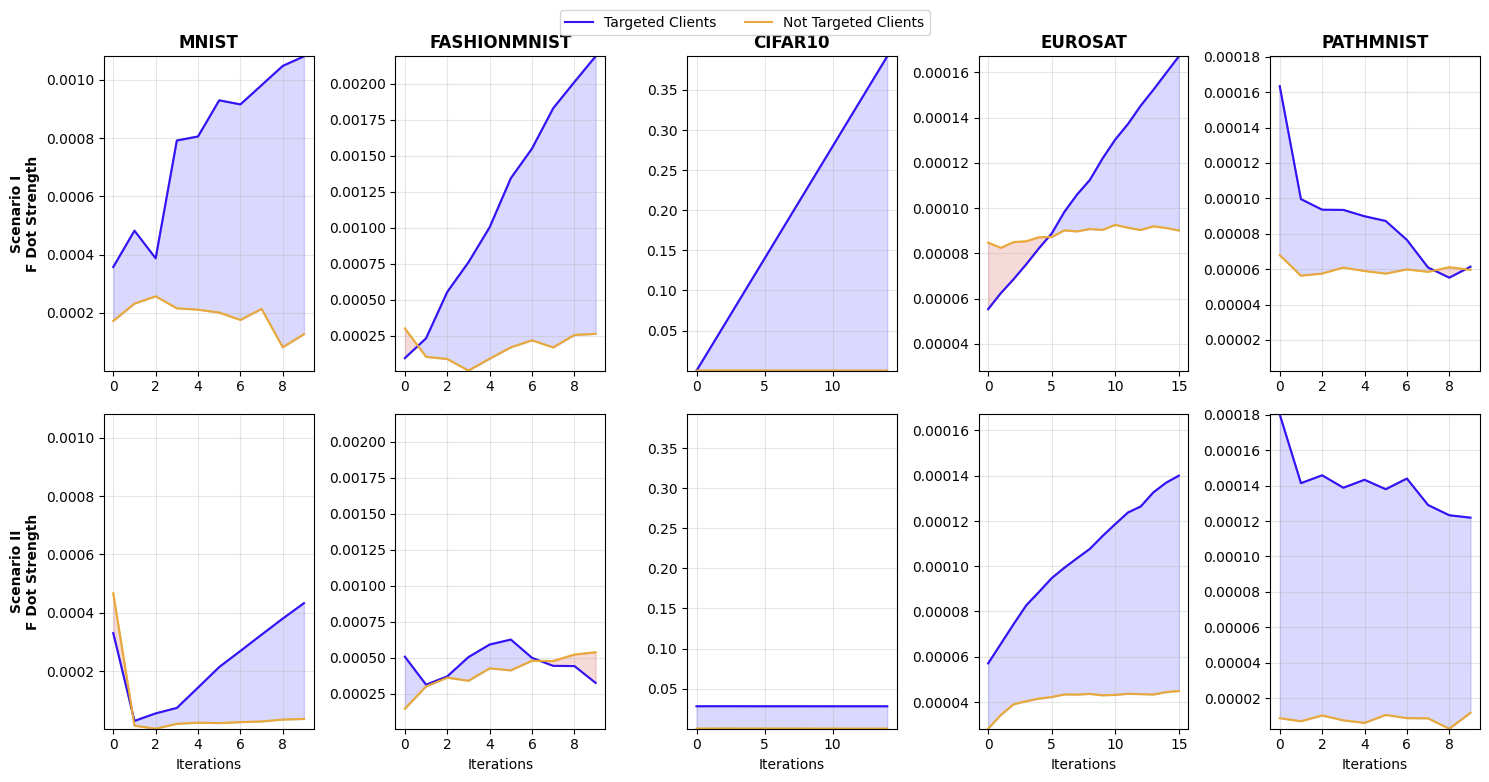}
   
    \caption{Fingerprint strength for FedAvg (20 clients) across all five datasets. Columns show Scenario I and II; rows show datasets. Blue areas indicate stronger fingerprints in targeted clients (expected); red areas indicate the opposite.}
  
    \label{fig:fingerprint_20_fedavg}
\end{figure*}

\paragraph{\textbf{Scalability analysis}}
To assess robustness in larger client populations, we conduct scalability experiments. We replaced CIFAR100 with FashionMNIST \cite{Xiao2017FashionMNISTAN} to reduce training cost while preserving class diversity. For MNIST, FashionMNIST, CIFAR-10, and PATHMNIST, we evaluate configurations with 20 and 50 clients; for EUROSAT, we use 10 and 20 clients due to dataset size constraints.
In \textbf{Scenario~I} (single targeted client), client~1 is targeted, and all other clients are honestly aggregated. One honest client is additionally designated as verified to monitor orchestrator behaviour and estimate false positives. In \textbf{Scenario~II} (multiple targeted clients), for 20 clients, two are targeted (clients~0–1) and two verified (clients~1–2); for 50 clients, four are targeted (clients~0–3) and four verified (clients~2–5). These settings enable evaluation of detection speed, reliability, and scalability across varying dataset sizes and client populations under non-IID distributions.

\subsection{Model structure and training parameters}

For MNIST and FashionMNIST~\cite{Xiao2017FashionMNISTAN}, we use a lightweight CNN composed of two $3 \times 3$ convolutional layers (padding~1), each followed by ReLU activation and $2 \times 2$ max pooling, a fully connected layer with 128 units, and a final linear classifier over 10 classes.
For CIFAR-10~\cite{krizhevsky2009Cifar10}, CIFAR-100~\cite{krizhevsky2009learning}, PATHMNIST~\cite{yang2023medmnist}, and EUROSAT~\cite{helber2019eurosat}, we adopt ResNet-18~\cite{Kaiming_ResNet}.

The number of local training epochs is set to 4 for simpler datasets (MNIST and PATHMNIST) and 6 for more complex datasets (CIFAR-10, CIFAR-100, and EUROSAT). The full reproducible code is available in \href{https://github.com/MKZuziak/PoisontoDetect/tree/main}{this GitHub repository}.
\subsection{Cross-silo evaluation}
\subsubsection{Detection using label flipping}\label{cross_sile_evaluation_flip}
\begin{figure*}
    \centering
    \begin{subfigure}[t]{1\linewidth}
        \centering
        \includegraphics[width=1\linewidth]{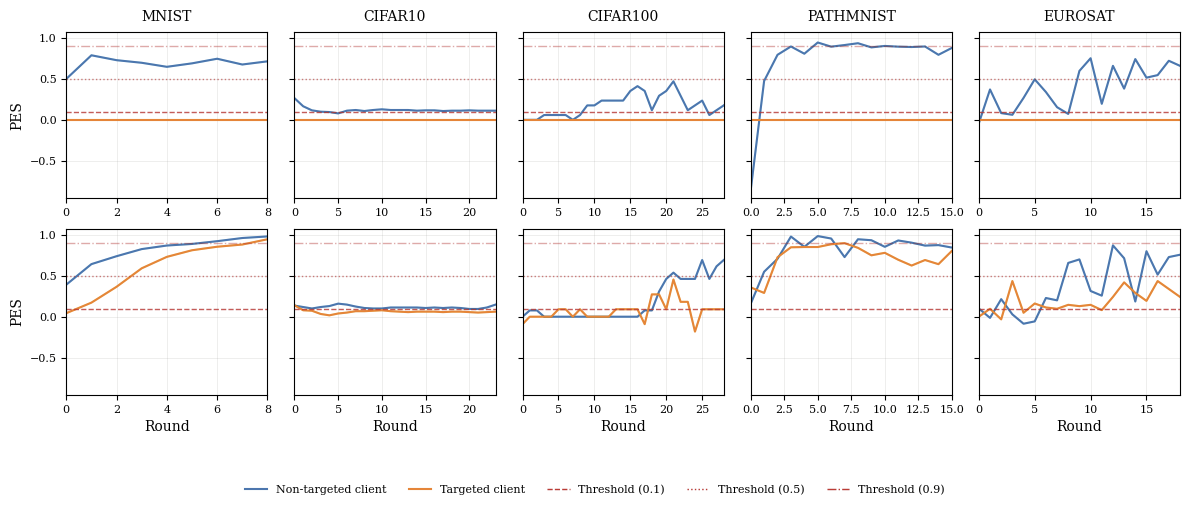}
       
        \caption{PES scores per round for the Label Flip (Algorithm~\ref{alg:label_flipping}) across all five datasets. Columns show Scenario I and II; rows show datasets. Dotted red lines indicate thresholds (0.1, 0.5, 0.9). Dishonestly aggregated models should exhibit lower PES than honestly aggregated ones.}
        
        \label{fig:label_flip_lineplot}
    \end{subfigure}
    
    \vspace{1em}
    
    \begin{subfigure}[t]{1\linewidth}
        \centering
        \includegraphics[width=1\linewidth]{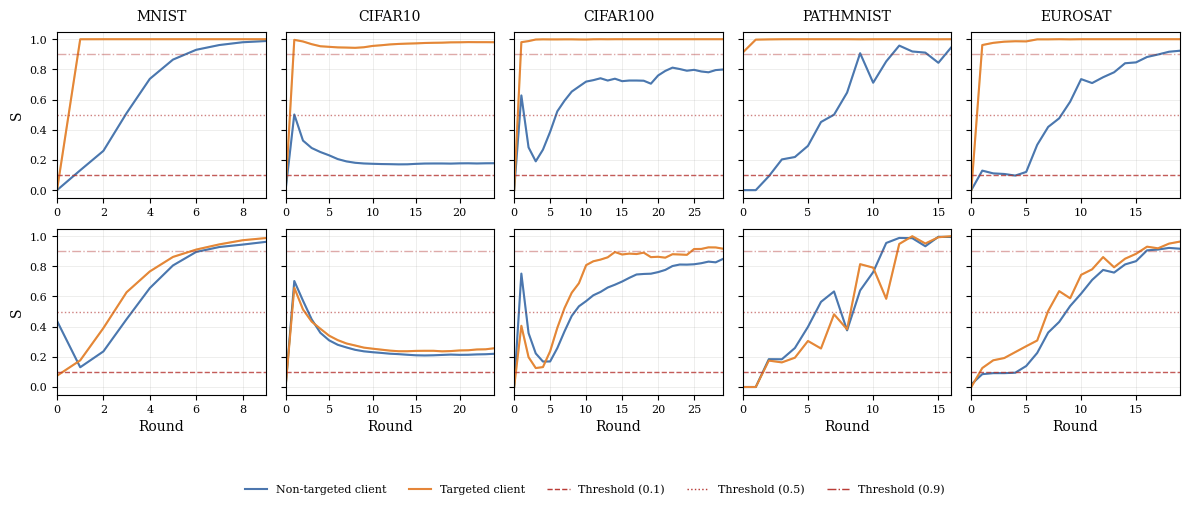}
       
        \caption{Backdoor strength $S$ scores per round for the Backdoor Trigger (Algorithm~\ref{alg:backdoor_trigger}) across all five  datasets. Columns show Scenario I and II; rows show datasets. Dotted red lines indicate thresholds (0.1, 0.5, 0.9). Dishonestly aggregated models should exhibit higher $S$ than honestly aggregated ones.}
        
        \label{fig:backdoor_lineplot}
    \end{subfigure}

    \caption{Detection metrics across rounds for (a) Label Flip using PES scores (Algorithm~\ref{alg:label_flipping}) and (b) Backdoor Trigger using strength $S$ (Algorithm~\ref{alg:backdoor_trigger}), across all five datasets and two scenarios.}
    \label{fig:poisoning_detection_comparison}
\end{figure*}

\begin{figure*}
    \centering
    \begin{subfigure}[t]{1\linewidth}
        \centering
        \includegraphics[width=1\linewidth,height=6cm]{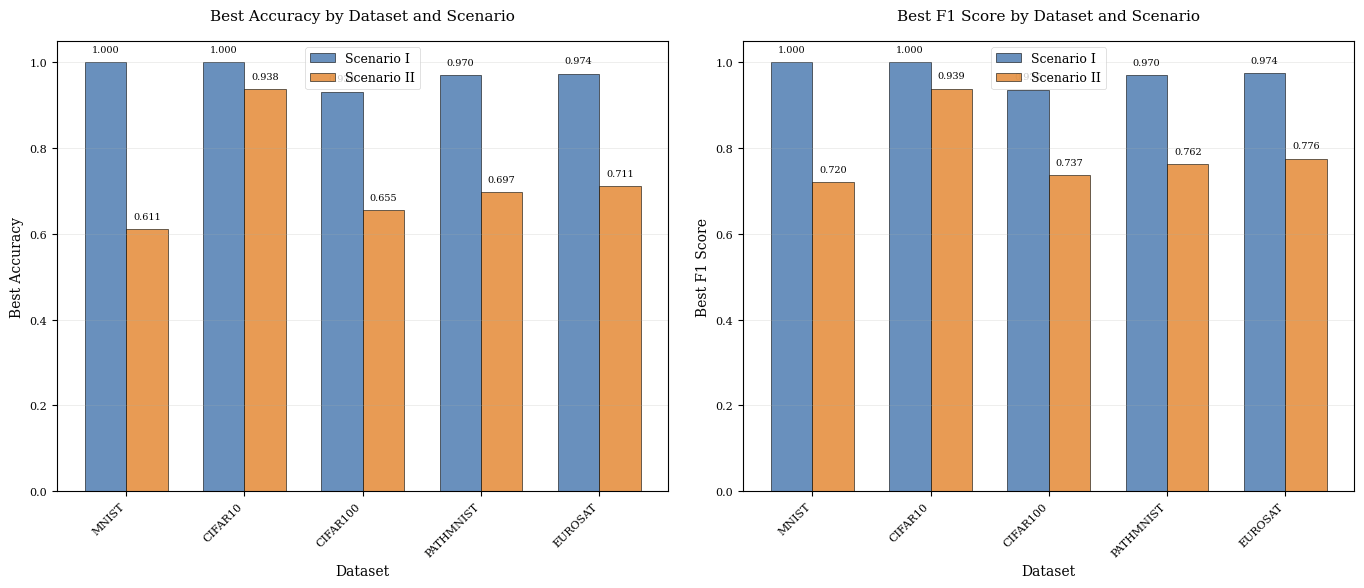}
        \caption{The best achievable results for the performance of the Label Flip (Algorithm \ref{alg:label_flipping}) on various datasets (Scenario I and Scenario II). The left-hand figure places each dataset (x-axis) against a best measured accuracy (y-axis), while the right-hand figure places the same datasets against a best measured F1 Score (y-axis). The blue bar presents the scores achieved in Scenario I, while the orange bar presents the scores achieved in Scenario II.}
        \label{fig:label_flip_best}
    \end{subfigure}
    
    \vspace{1em}
    
    \begin{subfigure}[t]{1\linewidth}
        \centering
        \includegraphics[width=1\linewidth,height=6cm]{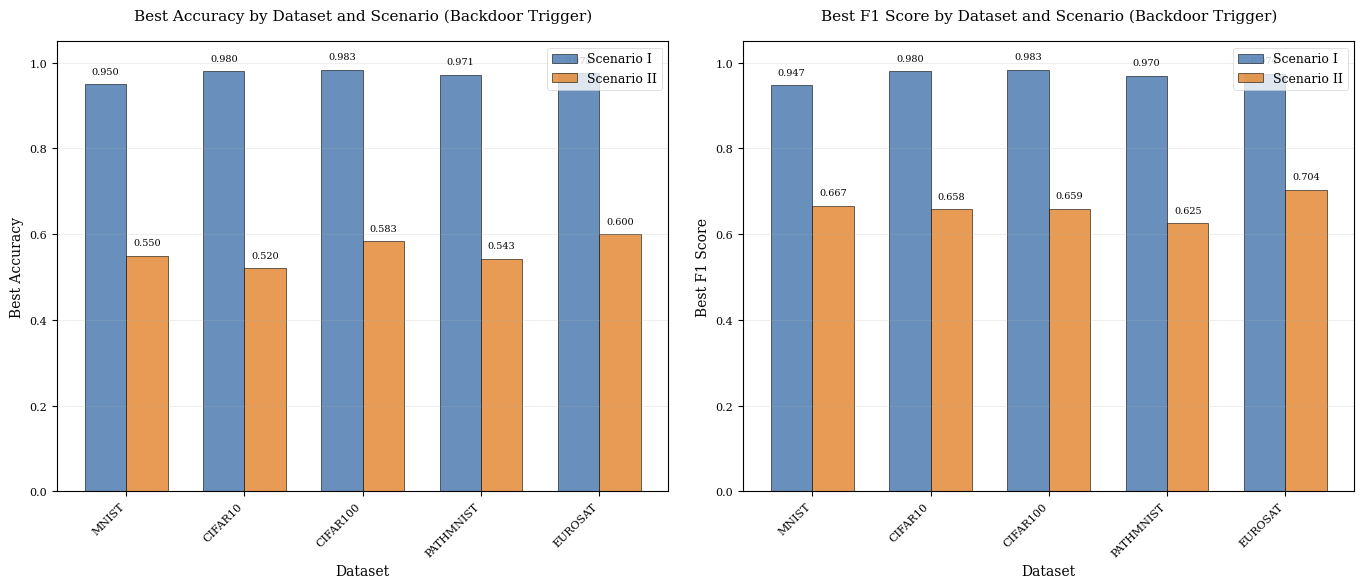}
        \caption{The best achievable results for the performance of the Backdoor Trigger (Algorithm \ref{alg:backdoor_trigger}) on various datasets (Scenario I and Scenario II). The left-hand figure places each dataset (x-axis) against a best measured accuracy (y-axis), while the right-hand figure places the same datasets against a best measured F1 Score (y-axis). The blue bar presents the scores achieved in Scenario I, while the orange bar presents the scores achieved in Scenario II.}
        \label{fig:backdoor_best}
    \end{subfigure}

    \caption{Best achievable detection performance for (a) Label Flip (Algorithm~\ref{alg:label_flipping}) and (b) Backdoor Trigger (Algorithm~\ref{alg:backdoor_trigger}) across datasets and scenarios, measured by Accuracy and F1 Score.}
    \label{fig:best_performance_comparison}
\end{figure*}

We evaluate the method on both targeted and non-targeted clients to account for false positives. 
Figure~\ref{fig:label_flip_lineplot} shows iteration-by-iteration PES scores, consistently higher for non-targeted clients. The difference is more pronounced in Scenario~I than Scenario~II, yet the method reliably detects orchestrator-induced targeted overfitting in both scenarios.
% =====ADED====
To quantify potential false positives, we monitored the PES score of a verified honest client in each scenario; scores remained consistently low across datasets, indicating rare misclassification of honest clients.
%============
Table~\ref{tab:comprehensive_method_comparison} shows detection accuracy between 0.7 and nearly 1.0 in Scenario~I (except CIFAR-10), and targeted overfitting is detected very early (as early as round~1) across most datasets and scenarios (Table~\ref{subtab:detection_rounds}). Figure~\ref{fig:label_flip_best} shows a slight drop in performance for Scenario~II, attributable to the dilution of the poisoned signal when multiple clients are targeted simultaneously, which makes subtle overfitting harder to distinguish. Overall, changing the number of aggregated clients can shift the algorithm’s decision boundary.

\paragraph{\textbf{Limitations}} The label flipping method requires multiple local epochs (typically 4–7) for the model to learn the flipped patterns, and heterogeneous (non-IID) data can weaken the poison signal due to uneven client convergence, reducing detection reliability. Under Scenario II, signal dilution across simultaneously targeted clients further reduces PES separability. Increasing $\alpha_{\text{flip}}$ or lowering $\tau_{\text{pes}}$ can partially compensate at the cost of a marginally higher utility impact and false positive rate, respectively. If inter-client communication is permitted, clients can coordinate by agreeing on a single shared class to poison, concentrating their detection signals in the malicious aggregate. It also computes accuracy on the poisoned subset twice per iteration—before sending the local model and after aggregation—making it computationally costly. Although manageable for small cross-silo experiments, this overhead scales linearly with the client number and local data size. The threshold $\tau_{pes}$ is easy to tune, bounded in $[0,1]$, with lower values increasing sensitivity.

\subsubsection{Detection using backdoor trigger}

Backdoor-based detection reduces runtime by nearly half compared to label flipping, since the backdoor strength is computed only after aggregation, while achieving similar detection performance.

Figure \ref{fig:backdoor_lineplot} illustrates the progression by iteration of the backdoor detection scores \textit{S} (Trigger Influence Score), where higher \textit{S} scores reflect a dishonest aggregation. Targeted clients consistently score higher than non-targeted clients. In Scenario~I, the decision boundary between cohorts is clear, while in Scenario~II it is more subtle, increasing the chance of false positives. Figure~\ref{fig:backdoor_threshold_s2} confirms this: best detection accuracy ranges 0.95–0.98 for Scenario~I and 0.52–0.6 for Scenario~II. Detection is almost immediate in Scenario~I (Table~\ref{subtab:detection_rounds}), but requires several rounds in Scenario~II when multiple clients are targeted. In general, the backdoor method exhibits a behaviour similar to that of label flipping.

\paragraph{\textbf{Limitations}} 
Uneven client convergence in non-IID settings can weaken the encoded signal, reducing the detection reliability. Increasing the size of the trigger sample or using distinctive trigger patterns could mitigate this effect.

Under Scenario II, the trigger signal is additionally diluted through averaging across simultaneously targeted clients, which reduces the Trigger Influence Score $\mathcal{S}$ and increases the likelihood of missed detection. This effect becomes more pronounced when targeted clients use independent trigger patterns, since the resulting signals may become misaligned in the aggregated update. If inter-client coordination is possible, clients can mitigate this issue by agreeing on a shared trigger pattern and target label, thereby reinforcing a common signal in the malicious aggregate.
When clients already share the same trigger configuration --- as in our experimental setup --- dilution arises primarily from the averaging process itself. In this setting, increasing 
$|\mathcal{D}_{\text{trigger}}|$ strengthens the trigger association in the aggregated model and partially compensates for the dilution effect, although at the cost of a small utility trade-off.

\subsubsection{Detection using fingerprinting} Fingerprinting achieves strong and robust performance across all datasets in Scenario~I (Tables~\ref{subtab:scenario_i_metrics}, \ref{subtab:detection_rounds}), with near-immediate detection and accuracy ranging from 0.75 to 1.0, making it the most reliable method in this setting. However, performance decreases significantly in Scenario~II, where a subset of clients is targeted. In this case, the fingerprint signal is diluted by aggregation with other targeted clients, requiring stronger fingerprints to remain detectable. Such stronger fingerprints interfere with training and degrade model performance, making the method unreliable in Scenario~II. 

We explore three fingerprint generation strategies: sparse fingerprints, which minimise training interference; random unit vectors, which offer increased robustness under aggregation; and hash-based fingerprints, which enable deterministic reproducibility and facilitate collaborative calibration. The choice of fingerprint affects the detection sensitivity, model performance, and scalability.
\begin{figure*}[ht]
    \centering

    \includegraphics[width=0.85\textwidth]{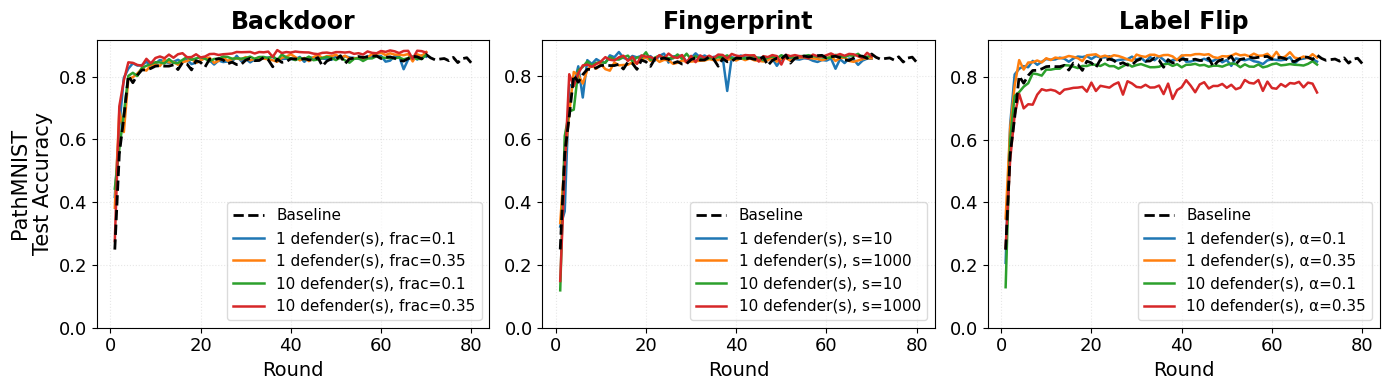}
    \caption{Orchestrator test accuracy per detection method on PATHMNIST (20 clients, FedAvg), baseline vs one and ten detecting clients under two parameter settings.}
    \label{fig:utility_pathmnist}
\end{figure*}
\paragraph{\textbf{Limitations}} 
The fingerprinting method assumes that the orchestrator’s optimiser both receives and returns the same type of update (either gradients or model weights). As a result, optimisers that receive gradients but return weights ( e.g., FedProx) are incompatible. Additionally, in Scenario~II, stronger fingerprints are required to counteract signal dilution, which can disrupt learning and degrade model performance.

Client-specific fingerprint vectors are constructed using independent client seeds, producing sparse random vectors that are approximately orthogonal in high-dimensional parameter spaces. Under independence assumptions, their inner products are zero in expectation, and concentration effects ensure that empirical inner products remain negligible with high probability.\footnote{This follows from the Johnson--Lindenstrauss lemma~\cite{johnson1984extensions}, which guarantees that random projections in sufficiently high-dimensional spaces preserve approximate orthogonality with high probability.}

Despite this approximate orthogonality, each client's fingerprint signal is diluted by a factor of approximately $1/n$ under uniform aggregation across $n$ clients simultaneously targeted. Without inter-client coordination, this effect can be mitigated by increasing the target dot strength $S_{\text{target}}$, which scales the fingerprint amplitude and partially compensates for dilution, at the cost of a small utility trade-off (see Section~\ref{sec:utility}).

If inter-client communication is permitted, clients can coordinate the construction of nearly orthogonal fingerprint directions $\{s_1, s_2, \ldots, s_n\}$, improving signal separability in the aggregated update, reducing cross-client interference, and providing a formal orthogonality guarantee.

\subsection{Comparison of the detection mechanisms}
Overall, our results highlight the trade-offs among detection mechanisms (see table \ref{subtab:detection_rounds} and table \ref{subtab:scenario_i_metrics}). In \textbf{Scenario~I}, fingerprinting is the fastest and most reliable method, achieving immediate detection with near-perfect scores. Label flipping follows closely, typically detecting the attack within the first rounds but with more moderate scores. In contrast, the backdoor trigger method exhibits higher variability across datasets. To analyse the sensitivity of label flipping and backdoor detection, we study the effect of detection thresholds $\tau_{pes}$ and $\theta$ on accuracy, precision, recall, and F1-score in nine threshold configurations. The results are shown in Figures~\ref{fig:labelflip_threshold_s1} and~\ref{fig:labelflip_threshold_s2} (label flip), and Figures~\ref{fig:backdoor_threshold_s1} and~\ref{fig:backdoor_threshold_s2} (backdoor trigger).

\textbf{Scenario~II} is inherently more challenging, as targeted aggregation involves only a subset of clients, reducing the separability of the signal. In this setting, label flipping achieves the most reliable performance, followed by the backdoor trigger method, while fingerprinting fails to reliably distinguish targeted overfitting when multiple clients are targeted.

\begin{table*}[t]
  \centering

  \footnotesize
  \renewcommand{\arraystretch}{1.2}
  
  % Detection Rounds Subtable
  \begin{subtable}{\textwidth}
    \centering
    \begin{tabular}{llcccccc}
      \toprule
      \multirow{2}{*}{\textbf{Scenario}} & \multirow{2}{*}{\textbf{Method}} &
        \multicolumn{5}{c}{\textbf{Datasets}} \\
      \cmidrule(lr){3-7}
      & & \textbf{MNIST} & \textbf{CIFAR10} & \textbf{CIFAR100} & \textbf{PATHMNIST} & \textbf{EUROSAT} \\
      \midrule
      \multirow{3}{*}{Scenario I}
        & Label Flip        & 1 & 1 & 1 & 1 & 1 \\
        & Backdoor Trigger  & 2 & 2 & 2 & 1 & 2 \\
        & Fingerprint       & 1 & 1 & 1 & 1 & 1 \\
      \midrule
      \multirow{3}{*}{Scenario II}
        & Label Flip        & 1 & 1 & 1 & 10 & 1 \\
        & Backdoor Trigger  & 3 & 2 & 2 & 6 & 6 \\
        & Fingerprint       & 8 & {--} & {--} & {--} & {--} \\
      \bottomrule
    \end{tabular}
     \caption{Detection rounds by method and scenario}
     \label{subtab:detection_rounds}
  \end{subtable}
  
  \vspace{1.5em}
  
  % Scenario I Performance Subtable
  \begin{subtable}{\textwidth}
    \centering
    \begin{tabular}{llccccc}
      \toprule
      \multirow{2}{*}{\textbf{Metric}} & \multirow{2}{*}{\textbf{Method}} &
        \multicolumn{5}{c}{\textbf{Scenario I Datasets}} \\
      \cmidrule(lr){3-7}
      & & \textbf{MNIST} & \textbf{CIFAR10} & \textbf{CIFAR100} & \textbf{PATHMNIST} & \textbf{EUROSAT} \\
      \midrule
      \multirow{3}{*}{Accuracy}
        & Label Flip        & 1.000 & 0.521 & 0.707 & 0.970 & 0.842 \\
        & Backdoor Trigger  & 0.550 & 0.900 & 0.517 & 0.629 & 0.625 \\
        & Fingerprint       & 0.750 & 1.000 & 1.000 & 1.000 & 1.000 \\
      \midrule
      \multirow{3}{*}{Precision}
        & Label Flip        & 1.000 & 0.511 & 0.630 & 0.941 & 0.760 \\
        & Backdoor Trigger  & 0.529 & 0.857 & 0.509 & 0.567 & 0.576 \\
        & Fingerprint       & 0.667 & 1.000 & 1.000 & 1.000 & 1.000 \\
      \midrule
      \multirow{3}{*}{Recall}
        & Label Flip        & 1.000 & 1.000 & 1.000 & 1.000 & 1.000 \\
        & Backdoor Trigger  & 0.900 & 0.960 & 0.967 & 1.000 & 0.950 \\
        & Fingerprint       & 1.000 & 1.000 & 1.000 & 1.000 & 1.000 \\
      \midrule
      \multirow{3}{*}{F1 Score}
        & Label Flip        & 1.000 & 0.676 & 0.773 & 0.970 & 0.864 \\
        & Backdoor Trigger  & 0.667 & 0.906 & 0.667 & 0.723 & 0.717 \\
        & Fingerprint       & 0.800 & 1.000 & 1.000 & 1.000 & 1.000 \\
      \bottomrule
    \end{tabular}
    \caption{Detection accuracy for Scenario I}
    \label{subtab:scenario_i_metrics}
  \end{subtable}
    \caption{Comprehensive comparison of poisoning detection methods across datasets and scenarios. Detection rounds and Scenario I performance metrics, in Label Flip ($\alpha < 0.01$), Backdoor Trigger ($\theta > 0.9$), and Fingerprint methods.}
    \label{tab:comprehensive_method_comparison}
\end{table*}

\section{Utility analysis}
\label{sec:utility}

We evaluate the utility impact of each detection method under honest aggregation. Figure~\ref{fig:utility_pathmnist} shows orchestrator test accuracy across rounds on PATHMNIST (20 clients, FedAvg), comparing the baseline (no detection) against one and ten simultaneously detecting clients under two parameter settings per method.

Across all methods, a single detecting client produces negligible degradation, with accuracy curves remaining indistinguishable from the baseline. With ten detecting clients, backdoor trigger detection remains negligible under both settings, while fingerprinting exhibits only a small transient dip at $\alpha_{\text{fp}} = 9.1 \times 10^{-5}$ with ten detectors that recover during training. Label flipping remains negligible at $\alpha_{\text{flip}} = 0.1$, but a persistent gap below the baseline appears at $\alpha_{\text{flip}} = 0.35$ with ten detectors.

These observations align with the bounds derived in Appendix~\ref{sec:utility_bounds}. For label flipping ($K_{\text{classes}} = 9$, $C = \mathcal{O}(1)$), the bound predicts the worst-case degradation of 
$0.056\% \cdot C$ and $0.556\% \cdot C$ for one and ten detectors at $\alpha_{\text{flip}} = 0.1$, increasing to $0.194\% \cdot C$ and $1.944\% \cdot C$ at $\alpha_{\text{flip}} = 0.35$, consistent with the observed $\sim 2\%$ persistent gap. For fingerprinting, the bound scales as $\mathcal{O}(\alpha_{\text{fp}}^2 f_{\text{det}}/N)$, evaluating to $\mathcal{O}(2.1 \times 10^{-10})$ at $\alpha_{\text{fp}} = 9.1 \times 10^{-5}$ and $\mathcal{O}(2.1 \times 10^{-6})$ at $\alpha_{\text{fp}} = 9.1 \times 10^{-3}$, implying negligible perturbation cost across detector counts. The transient dip, therefore, likely arises from the client drift term $\mathcal{O}(\eta^2 E^2 \Delta_{\text{grad}})$ rather than the fingerprint perturbation itself. For backdoor trigger detection, no distribution-independent accuracy bound is available (see Appendix~\ref{sec:utility_bounds}), although empirical results show negligible degradation in both settings on PATHMNIST.

Additional results on MNIST and EUROSAT are in Appendix~\ref{sec:utility_additional}.
\begin{figure}[t]
    \centering
    \includegraphics[width=0.85\linewidth,height=4cm]{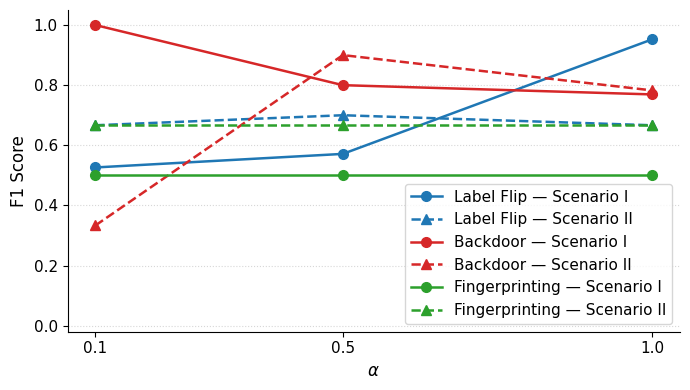}
    \caption{Detection F1 Score (first 10 rounds) by Dirichlet $\alpha$ on MNIST for all three methods, using the best-performing threshold per $\alpha$ (see Table~\ref{tab:noniid-thresholds}).}
   
    \label{fig:mnist-alpha}
\end{figure}
\section{Impact of heterogeneity}
We evaluate how data heterogeneity, controlled through the Dirichlet concentration parameter $\alpha$, affects detection reliability across all three methods. We report this effect in Figure \ref{fig:mnist-alpha} across all three methods on MNIST (20 clients, FedAvg) using fixed thresholds across all settings. We also report additional results in Appendix \ref{appendix:non_iid_additional}.
The detection reliability under varying $\alpha$ differs across methods 
and scenarios, depending on the balance between overfitting severity 
and client drift, both of which increase at low $\alpha$.
Label-flipping Scenario I consistently improves as $\alpha$ increases  (F1: $0.84 \rightarrow 1.0 \rightarrow 1.0$) because a reduced drift yields a cleaner PES signal. In Scenario II, F1 decreases slightly at higher $\alpha$ because the malicious aggregate over similarly distributed targeted clients partially corrects the poisoned signal, reducing the PES gap between targeted and non-targeted clients. Backdoor trigger Scenario I degrades with $\alpha$ under fixed thresholds, since the trigger signal weakens as overfitting becomes less severe; in Scenario II, detection nearly collapses at $\alpha = 0.1$ (F1 $= 0.20$) due to severe trigger signal misalignment across simultaneously targeted clients. Fingerprinting Scenario I remains flat across all $\alpha$ (F1 $\approx 0.50$), suggesting that detection is limited by threshold calibration at this cohort size rather than by data heterogeneity. In Scenario II, F1 improves with $\alpha$ (F1: $0.66 \rightarrow 0.95 \rightarrow 0.95$) because near-IID distributions reduce gradient misalignment among simultaneously targeted clients, thereby strengthening the aggregated fingerprint signal.
In order to address this challenge without inter-client communication, clients can partially mitigate this by normalising raw detection metrics (PES, trigger influence score, fingerprint strength) by their local variance observed in recent rounds, adapting naturally to non-IID-induced fluctuations. 
If inter-client communication is permitted, approximate Private Set Intersection (fuzzy-PSI)\cite{blass2025fuzzypsi, vanbaarsen2024fuzzy} can be used to identify peers with similar distributions. Clients compute local distribution descriptors(\eg discretised label frequencies) and identify peers within a similarity threshold without revealing raw data, and then jointly calibrate detection thresholds using pooled honest-round observations.

\section{Scalability evaluation}
We extend the cross-silo evaluation to medium-sized federated cohorts to assess how detection signals behave as the number of participating clients increases. Increasing the cohort size amplifies the noise in aggregation, making it more difficult for the verification signals to remain distinguishable. Based on the results of the cross-silo experiments, we focus on label flipping and gradient fingerprinting, as these methods are more suitable for this setting. Backdoor-based verification is not considered further, as it typically requires more local epochs to reliably encode the poison signal. Dataset descriptions and data splits are identical to those presented in Section~\ref{dataset_description}. We evaluate both methods with 20 and 50 clients.

\subsection{Label flip evaluation}
We evaluate the label flip method in 20- and 50-client cohorts to track the behaviour of targeted and non-targeted clients throughout training. FedAvg results for 20 clients are reported in the main text; FedOpt and 50-client results are provided in Appendix~\ref{appendix:scalability_expanded}. Detection is based on PES scores: when the dispatched and received models are similar (PES near baseline 0.0), the target overfitting is indicated. Consequently, a higher area under the threshold corresponds to a lower probability that the client is attacked. The method reliably identifies targeted overfitting, with cohort separation decreasing as client numbers increase, raising false positives in Scenario~II.

\subsection{Fingerprinting evaluation}
We examine the persistence of fingerprint signatures in targeted and non-targeted clients under aggregation for 20- and 50-client cohorts. The fingerprint scaling parameter $\alpha$ (Algorithm~\ref{alg:gradient_fingerprinting}) controls signal strength: appropriately chosen, it ensures that targeted client fingerprints prevail post-aggregation, while non-targeted fingerprints diminish. FedAvg results for 20 clients are shown in Figure~\ref{fig:fingerprint_20_fedavg}. The figures indicate whether the fingerprints of the targeted clients exceed those of the non-targeted clients: blue (positive) areas indicate stronger fingerprints in the targeted clients, whereas red (negative) areas indicate the opposite. Ideally, the area is positive, showing fingerprints persist in targeted aggregation and diminish in non-targeted clients. As the cohort size or complexity of the optimiser increases, this separation becomes more sensitive to $\alpha$, although the method remains effective when $\alpha$ is properly tuned.

\subsection{Scalability implications}
Scalability experiments show that detection effectiveness depends on both cohort composition and method parameters. Label flipping remains reliable for detecting targeted overfitting in Scenario I, but false positives increase in Scenario II, requiring careful threshold selection. Fingerprinting is generally more robust, maintaining separation between targeted and non-targeted clients across cohort sizes and optimiser types, although larger cohorts or more complex optimisers reduce signal visibility, emphasising the need to calibrate the fingerprint strength $\alpha$. Overall, scalability is governed less by client count alone than by how detection signals survive aggregation under optimiser-induced noise.

\begin{figure*}[t]
    \centering
    \includegraphics[width=0.85\linewidth,height=13cm]{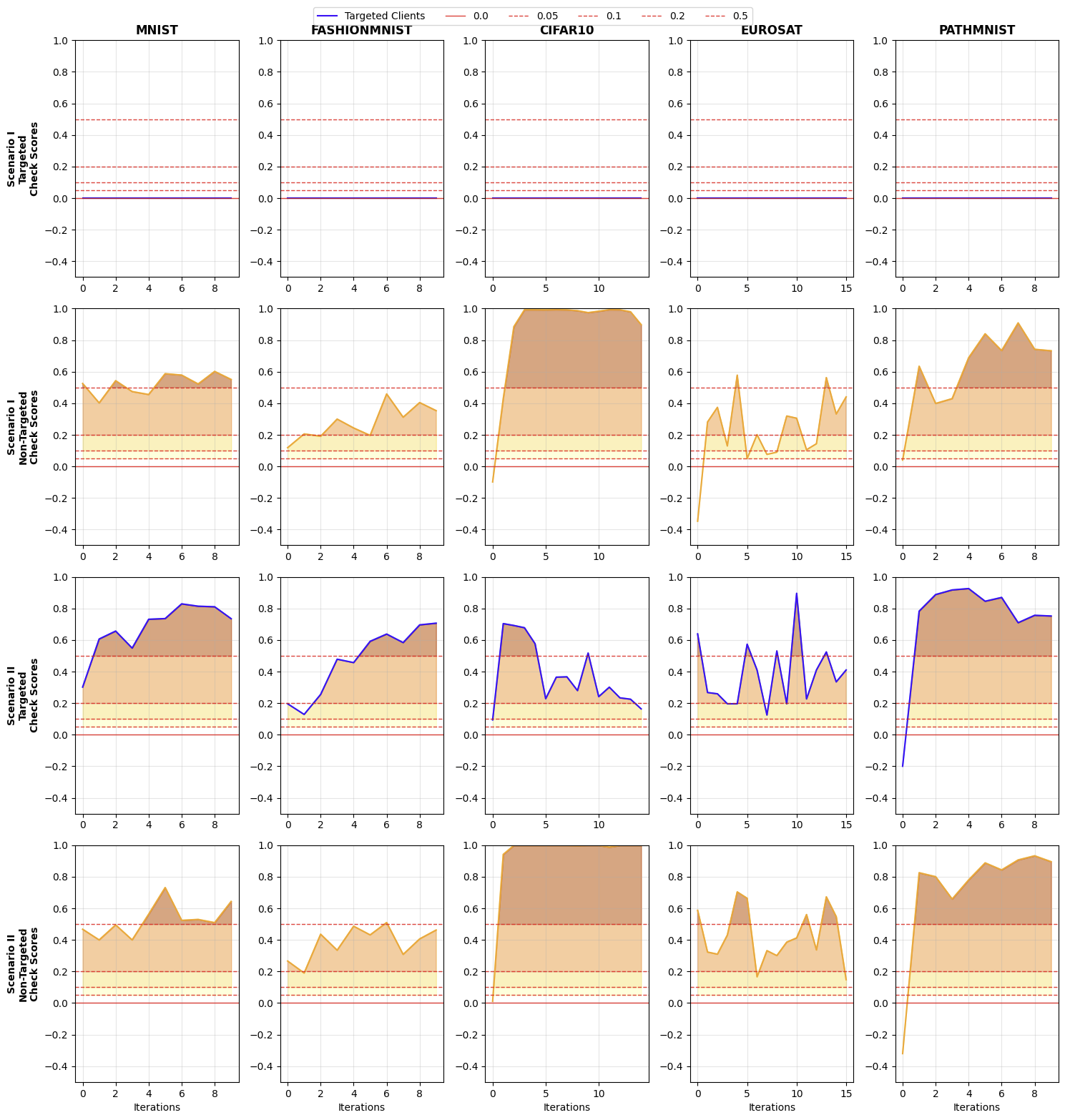}
   
    \caption{PES scores for Label Flip (FedAvg, 20 clients) across all five datasets. Rows show targeted and non-targeted clients for Scenario I (rows 1--2) and Scenario II (rows 3--4); columns show datasets. Dotted red lines indicate thresholds (0.0, 0.05, 0.1, 0.2, 0.5). Non-targeted clients are expected to exceed the thresholds; targeted clients are expected to remain below them.}
   
    \label{fig:label_flip_20_fedavg}
\end{figure*}

\section{Discussion}

\paragraph{\textbf{Utility versus detection}}
Detection is entirely at the client's discretion. 
A client may choose to accept a small, controlled utility cost in exchange for a detection capability ---\eg at most $\sim 2\%$ accuracy degradation under label flipping at $\alpha_{\text{flip}} = 0.35$, $f_{\text{det}} = 0.5$, and $K_{\text{classes}} = 9$ (see Section~\ref{sec:utility}), similar in spirit to computational overhead accepted in security protocols. The cost is controlled through method-specific parameters: $\alpha_{\text{flip}}$ and the detection frequency for label flipping, $\alpha_{\text{fp}}$ for fingerprinting, and $|\mathcal{D}_{\text{trigger},i}|/|\mathcal{D}_i|$ for backdoor trigger detection. As formally analysed in Appendix~\ref{sec:utility_bounds}, this cost is bounded under the stated assumptions and, in the case of fingerprinting, decreases with cohort size. Unlike passive defences such as Differential Privacy, which impose a permanent utility-privacy tradeoff regardless of whether an attack is present, our detection mechanisms incur cost only during active participation in the detection procedure, and clients retain full control over whether and when to deploy them.

\paragraph{\textbf{False Positives in Non-IID Settings}}
In FL with highly non-IID data, a client whose local data distribution is unique may observe that the global model update closely matches its own local update. This happens because there is no “counter-weight” from other clients with similar data. As a result, detection metrics such as fingerprint strength or label-flipping scores can temporarily spike, making the client suspect targeted overfitting even though the aggregator is honest. True malicious looping, on the contrary, produces a rapid spike in the metric, allowing clients to distinguish between natural convergence and targeted overfitting.
\paragraph{\textbf{False positives under robust aggregations}} 

Label flipping and backdoor-based detection are essentially forms of controlled poisoning. If the orchestrator is ”Honest-but-Rigid”—meaning it uses robust aggregation such as Krum\cite{Kram_2017} or Trimmed-Mean\cite{timmed_mean}- it may interpret these modifications as malicious, leading to benign clients being flagged or excluded. In future work, we plan to investigate steganographic fingerprinting, where detection signals are constrained to remain within the natural variance of benign updates. Such fingerprints could reduce the risk of triggering robust aggregation defences while still allowing clients to detect targeted overfitting after aggregation.

\paragraph{\textbf{Compatibility with cryptographic schemes}}
The major cryptographic schemes employed in FL --- 
secure aggregation\cite{bonawitz2017practical}, homomorphic encryption~\cite{paillier1999public,gentry2009fully,aono2017privacy, zhang2020batchcrypt}, Secure Multi-Party Computation \cite{mohassel2017secureml}, and functional 
encryption~\cite{SUN2021_FE, GUAN2024_FE} --- share a common property: the server cannot observe clients' local models in 
plaintext, limiting its ability to mount downstream attacks 
such as membership inference or data reconstruction. However, preventing plaintext access is not equivalent to preventing 
targeted overfitting. Under SMPC-based secure aggregation, each client's update is masked with pairwise random seeds that cancel only across the full participant set --- any attempt to aggregate a subset yields an un-decryptable or invalid aggregate under the protocol. Under HE-based aggregation, selective aggregation remains computable on ciphertexts, leaving targeted overfitting intact; the orchestrator, however, cannot exploit the overfitted model in plaintext, thereby reducing the exposure of intermediate updates used in reconstruction attacks. Under functional encryption~\cite{SUN2021_FE, GUAN2024_FE}, the feasibility of targeted overfitting depends not only on the permitted function family but also on whether the orchestration protocol restricts selective client aggregation. If the FE-based aggregation protocol permits only aggregation over a fixed or sufficiently large client set, the orchestrator's ability to isolate targeted client updates is reduced; however, if the protocol permits arbitrary subset aggregation, targeted overfitting may still be achievable, even without access to individual decrypted gradients. Since targeted overfitting via subset aggregation is not feasible under SMPC-based secure aggregation, detection is not required for this specific attack vector. 

Under HE-based and FE-based approaches, where clients decrypt the global model locally, our detection methods remain applicable. In fully-encrypted pipelines where no decryption occurs, client-side detection is not applicable.

\paragraph{\textbf{Effects of different detection thresholds}}
Clients with varying sensitivity thresholds and varying detection methods may behave differently: those with stricter thresholds may be more likely to isolate themselves from training, while more relaxed clients remain. Diverse client-side thresholds make it difficult for the aggregator to eliminate all detection signals, increasing the robustness of client-side verification across the cohort.

\vspace{-1em}
\section{Conclusion}
In this article, we studied \textit{targeted overfitting}, a threat in which a malicious orchestrator induces overfitting in specific clients through selective aggregation, thereby exposing them to various attacks, such as membership inference and model inversion. To address this, we proposed three client-side detection mechanisms — label flipping, backdoor triggers, and gradient/weights fingerprinting — enabling autonomous detection without inter-client communication. Our evaluation showed that the three methods can successfully detect the threat when a single client is targeted, the fingerprinting method being the most reliable. However, when the attack targets multiple clients simultaneously, only the label flipping and backdoor trigger methods maintain effectiveness, whereas the fingerprinting method exhibits reduced performance. %We made our implementation and experimental data publicly available.\footnote{\href{https://anonymous.4open.science/r/Detect_targeted_overfitting-E646/README.md}{Github anonymised repository}}

In future works, we aim to study this problem from two distinct angles. Firstly, we would like to amplify the capabilities of the attacker (the orchestrator) by either incorporating more nuanced schemas of aggregation or different types of optimisers. Secondly, we would like to further examine the robustness of the fingerprinting method, which has a clear potential as a validation method, but requires further study in order to better understand the relevance of the incorporated fingerprint under various aggregation schemas.
%\section{Acknowledgments}
\begin{acks}
%For El Mestari and Lenzini, the work was supported by the following grants: []
For Zuziak, the following research was conducted, and the paper was drafted, during his stay at the National Research Council (CNR). %, and %was supported by the following grants: []. 
The research was published during his employment at the University of Leeds.
We would like to express our sincere gratitude towards the Interdisciplinary Centre for Security, Reliability and Trust (SnT) for access to high-performance computing resources granted to us for the sake of this research.
The authors used generative AI-based tools to revise the text, improve flow, and correct typos, grammatical errors, and awkward phrasing.
\end{acks}

\bibliographystyle{ACM-Reference-Format}
\bibliography{sample-base}

\appendix

\section{Appendix Overview}
This Appendix contains additional results and theoretical derivations that complement the findings presented in the main body. Section \ref{Appendix_prevention} contains ethical considerations for the disclosure of findings. Section \ref{Appendix_Memorisation} presents additional results on client-specific memorisation amplification. Section~\ref{sec:utility_bounds} derives worst-case upper bounds on the utility cost of each detection method under honest aggregation. Section~\ref{sec:utility_additional} reports utility analysis results on MNIST and EUROSAT, complementing the PATHMNIST results presented in Section~\ref{sec:utility}. Section \ref{Appendix_Best_Performing} reports the best achieved accuracy and F1 Score together with a detection threshold $\tau_{pes}$ used for achieving that score. Section \ref{Appendix_Threshold_Impact} presents additional results on the impact of various detection thresholds on performance across all five datasets. 

Section~\ref{appendix:non_iid_additional} reports detection performance across the Dirichlet concentration parameters $\alpha \in \{0.1, 0.5, 1.0\}$ for EUROSAT, examining how data heterogeneity affects detection reliability, while Section~\ref{appendix:scalability_expanded} presents additional results on the scalability of the presented solutions. 

\section{Proactive prevention of harm} \label{Appendix_prevention}
The following paper describes a possible attack in the Federated Learning scenario and provides an extensive evaluation of defence methods that clients can employ to detect the attacker in time, potentially preventing harm at the earliest stage. After careful consideration of the ethical implications, the authors conclude that the paper should pass the proportionality test, given that the benefits outweigh identifiable harms (if any).

The ethical evaluation was carried out using the \textit{Stakeholder-based Ethics Analysis} and was divided into (a) identification of relevant Stakeholders, (b) projecting the possible impacts, (c) implementing suitable mitigations against adversarial impacts (if any) and (d) final decisions regarding the disclosure of the outcomes presented in this paper.

The authors have adopted a broad definition of stakeholders, encompassing all entities that may be (negatively) influenced by the disclosure of the paper, including natural persons, organisations, and operating teams. Regarding those stakeholders, the impact analysis was carried out. As the article focuses more on defensive mechanisms than on introducing novel attack methods, it has been concluded that the disclosure in this paper can increase stakeholders' safety without posing any adverse effects on its members. The authors have sought to identify two types of possible (adversarial) impacts: infringement of ethical principles and tangible and intangible harms to stakeholders. In both cases, the authors concluded that the paper, in its current form and with its current artefacts, poses no danger to either aspect.

When presenting a novel defence mechanism, one often must pose the following question: \textit{Can these protective measures be used to harm others}? The authors have taken all necessary precautions to ensure that this will not occur with the disclosed methods. All the attacks were carried out in a simulated environment, without posing any threat to external sources or violating any infrastructure. Moreover, since the code was used only to simulate the attack (not to perform the real attack in the federated scenario), we note that the disclosed source material can not be used in any malicious way to harm third parties.

Based on the following analysis, we concluded that the source material can be disclosed and shared with the community without endangering the stakeholders. We sincerely believe that the detection methods described in this paper can enhance community safety, including for those who share their personal data under any federated aggregation schema. We also sincerely hope to draw the community's attention to the problem of \textit{false aggregation} detection through the angle proposed in this article. It must be noted that the relationship between model overfitting and the susceptibility to privacy-related attacks has long been known to the community. Yet, we feel that the problem is still not visible enough in the community, and too often the efficiency of the learning (federated) algorithm is used to justify the sacrifice of participants' privacy. We sincerely believe that discussion should not only be directed toward the question \textit{how can I protect myself from the attack}, but also \textit{whether I was attacked in the first place}. While the community has some answers to the former, it has not yet fully acknowledged the necessity of addressing the latter.
\section{Client-specific memorisation amplification}\label{Appendix_Memorisation}

\begin{figure}
    \includegraphics[width=\linewidth]{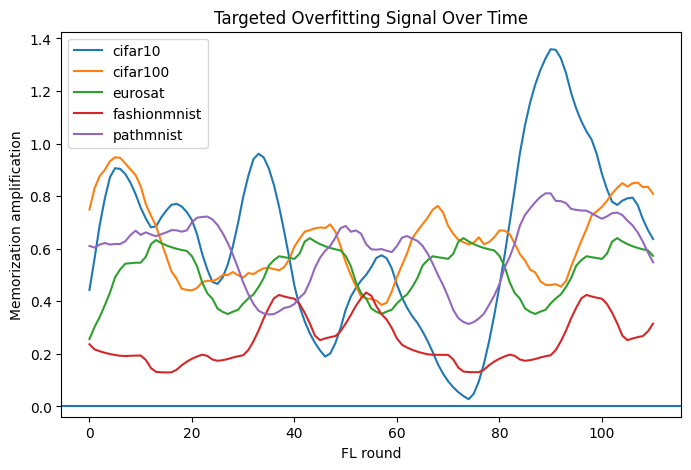}
    \caption{Memorisation amplification across federated learning rounds as defined in Equation \ref{eq:amplification} Blue vertical line placed at the $0.0$ threshold.}
    \label{fig:amplification}
\end{figure}
This section demonstrates that targeted overfitting produces client-specific memorisation that measurably increases data leakage risk.

For each federated round $t$, we define the target client generalisation gap as
\begin{equation}
\mathrm{Gap}_{\text{target}}^{(t)}
=
\mathcal{L}_{\text{ref}}^{(t)}
-
\mathcal{L}_{\text{target}}^{(t)},
\end{equation}
where $\mathcal{L}_{\text{target}}^{(t)}$ is the loss of the global model on data from the targeted client, and $\mathcal{L}_{\text{ref}}^{(t)}$ is the loss on data from non-targeted clients.

This quantity measures how well the model fits the targeted client relative to other client distributions.

 To determine whether memorisation is preferentially allocated to targeted clients, we compare the targeted gap to the corresponding gap computed for benign clients:
\begin{equation}
\mathrm{Amplification}^{(t)}
=
\mathrm{Gap}_{\text{target}}^{(t)}
-
\mathrm{Gap}_{\text{benign}}^{(t)}.
\label{eq:amplification}
\end{equation}

Positive amplification indicates that the global model fits the targeted client data more strongly than non-targeted clients' data relative to a common reference distribution, a pattern consistent with memorisation as characterised by prior work linking overfitting, generalisation gap, and information leakage from gradients \cite{Zhang_2017_generalisation},\cite{Fledman_2020}, \cite{zhu2019deep}, \cite{yin2021see}, \cite{carlini2023extracting}. Consequently, an increase in the target client generalisation gap reflects stronger encoding of targeted data in model parameters and, therefore, higher privacy exposure relative to the non-targeted clients.

Figures~\ref{fig:amplification} plots $Amplification^{(t)}$ across federated rounds, where persistent positive values indicate preferential memorisation of targeted client patterns compared to benign clients.
\FloatBarrier
\section{Bounding the utility cost of detection}
\label{sec:utility_bounds}

In this appendix, we derive worst-case upper bounds on the degradation of clean accuracy (or, where appropriate, a smooth surrogate objective) induced by each detection method under FedAvg. Assuming honest aggregation — the setting in which clients run detection precautionarily, but the orchestrator is, in fact, honest. We emphasise that the notion of ''utility cost'' is unified at the level of interpretation, while the specific measurable proxy depends on the analytical tractability of each mechanism.

The following analyses are conducted under a common set of assumptions. We assume that the loss function $F$ is $L$-smooth with $L = \mathcal{O}(1)$, and that clients perform $E$ local gradient steps with learning rate $\eta$ before 
aggregation. Dataset sizes across clients may be heterogeneous in general; simplifications to the equal-dataset-size case ($|\mathcal{D}_i| = \bar{D}, \ \forall i$) are explicitly stated where applicable. Gradient dissimilarity across clients is captured by:
\begin{equation}
\Delta_{\text{grad}} = \mathbb{E}_i\!\left[\|\nabla F_i(w) - \nabla F(w)\|^2\right],
\end{equation}
which depends on data heterogeneity and is generally not directly observable.
We additionally assume that each client runs exactly one detection 
method. All results are derived under weighted FedAvg aggregation.

\paragraph{Notation.}
Let $N$ denote the total number of clients and 
$\mathcal{S}_{\text{det}}$ the set of detecting clients, and Let $|\mathcal{D}_{\text{total}}| = \sum_{i=1}^N |\mathcal{D}_i|$. We define 
the dataset-size-weighted detecting fraction as:
\begin{equation}
f_{\text{det}} =
\sum_{i \in \mathcal{S}_{\text{det}}}
\frac{|\mathcal{D}_i|}{|\mathcal{D}_{\text{total}}|}.
\end{equation}
We also define the count-based detecting fraction as
$f_{\text{det}}^{(c)} = |\mathcal{S}_{\text{det}}|$, so that under 
equal dataset sizes ($|\mathcal{D}_i| = \bar{D}$ for all $i$),
\begin{equation}
f_{\text{det}} = \frac{f_{\text{det}}^{(c)}}{N}.
\end{equation}
Further, $K_{\text{classes}}$ denotes the number of classes, 
$\alpha_{\text{flip}}$ the label flip ratio, and 
$\alpha_{\text{fp}}$ the fingerprint strength scalar. For the 
backdoor trigger method, $|\mathcal{D}_{\text{trigger},i}|/|\mathcal{D}_i|$ 
denotes the trigger injection ratio for client $i$.

\subsection{Label flipping}
\label{sec:bound_labelflip}

When using label flipping, each detecting client independently selects its least frequent 
class $c_i^*$ in its local dataset $\mathcal{D}_i$, which 
satisfies:
\begin{equation}
\frac{|c_i^*|}{|\mathcal{D}_i|} \le \frac{1}{K_{\text{classes}}}.
\end{equation}
A fraction $\alpha_{\text{flip}}$ of samples in $c_i^*$ is 
flipped using the cyclic mapping $y \mapsto (y+1) \bmod 
K_{\text{classes}}$, producing a corrupted fraction per client $\beta_i$:
\begin{equation}
\beta_i = \alpha_{\text{flip}} \cdot 
\frac{|c_i^*|}{|\mathcal{D}_i|}
\le \frac{\alpha_{\text{flip}}}{K_{\text{classes}}}.
\end{equation}
Under FedAvg aggregation, the global corrupted fraction is:
\begin{equation}
\beta_{\text{global}} =
\sum_{i \in \mathcal{S}_{\text{det}}}
\frac{|\mathcal{D}_i|}{|\mathcal{D}_{\text{total}}|} \beta_i,
\end{equation}
which implies:
\begin{equation}
\beta_{\text{global}} \le
\frac{\alpha_{\text{flip}}}{K_{\text{classes}}}
\sum_{i \in \mathcal{S}_{\text{det}}}
\frac{|\mathcal{D}_i|}{|\mathcal{D}_{\text{total}}|}.
\end{equation}

Then:
\begin{equation}
\beta_{\text{global}} \le
\frac{f_{\text{det}} \cdot \alpha_{\text{flip}}}
{K_{\text{classes}}}.
\end{equation}
\paragraph{Client drift under FedAvg.}
Each client performs $E \in [4,7]$ local epochs before aggregation, leading to non-negligible client drift under non-IID data distributions.  Following standard federated optimisation analyses~\cite{mcmahan2017communication, karimireddy2020scaffold}, the deviation between FedAvg updates and centralised gradient descent scales as:
\begin{equation}
\epsilon_{\text{drift}} =
\mathcal{O}(\eta^2 E^2 \Delta_{\text{grad}}),
\end{equation}
where $\eta$ is the learning rate.

\paragraph{Accuracy degradation.}
Under bounded label noise assumptions and Lipschitz-smooth loss functions, the sensitivity of accuracy to corruption is bounded by a constant $C = \mathcal{O}(1)$. The overall degradation satisfies:
\begin{equation}
\Delta_{\text{Acc}}^{\text{lf}} \le
C \cdot \frac{f_{\text{det}} \cdot \alpha_{\text{flip}}}
{K_{\text{classes}}}
+ \mathcal{O}(\eta^2 E^2 \Delta_{\text{grad}}).
\end{equation}
When $\Delta_{\text{grad}}$ is small (near-IID, large Dirichlet $\alpha$), the drift term is negligible, and degradation is dominated by the label corruption term. When $\Delta_{\text{grad}}$ is large (high heterogeneity, small Dirichlet $\alpha$), the drift term dominates, and the effect of label flipping becomes secondary. Under example settings ($f_{\text{det}} = 1.0$, $\alpha_{\text{flip}} = 0.1$, $K_{\text{classes}} = 10$, $\eta = 0.005$, $E = 4$), the bound becomes:
\begin{equation}
\Delta_{\text{Acc}}^{\text{lf}} \leq 
0.01 \cdot C +
\mathcal{O}(4 \times 10^{-4} \cdot \Delta_{\text{grad}}).
\end{equation}
Here, $C = \mathcal{O}(1)$ is not instantiated numerically, as it depends on loss smoothness and hypothesis class complexity, and cannot be fixed without additional modelling assumptions.

\subsection{Backdoor trigger detection}
\label{sec:bound_trigger}

When using backdoor trigger detection, each detecting client injects a trigger pattern $T$ into a subset $\mathcal{D}_{\text{trigger},i} \subseteq \mathcal{D}_i$ with target label $y_\tau$, yielding a trigger injection ratio:
\begin{equation}
\beta_i =
\frac{|\mathcal{D}_{\text{trigger},i}|}{|\mathcal{D}_i|},
\end{equation}
and a global trigger injection fraction under FedAvg:
\begin{equation}
\beta_{\text{global}} =
\sum_{i \in \mathcal{S}_{\text{det}}}
\frac{|\mathcal{D}_i|}{|\mathcal{D}_{\text{total}}|}
\beta_i.
\end{equation}
Unlike random label noise, trigger poisoning introduces a structured and localised modification to the input space. In overparameterized models, such modifications may be partially decoupled from the representations that govern clean classification, depending on model capacity and feature overlap. As a result, the impact of trigger injection on clean accuracy is not necessarily monotonic or proportional to $\beta_{\text{global}}$.

Consequently, no general bound of the form $\Delta_{\text{Acc}} < f(\beta_{\text{global}})$ can be established in a distribution-independent manner without additional restrictive assumptions on model capacity, optimisation dynamics, or feature geometry.

\subsection{Gradient/weight fingerprinting}
\label{sec:bound_fingerprint}

We analyse the effect of gradient/weight fingerprinting under FedAvg by characterising the magnitude of the induced perturbation in the aggregated model update. Each detecting client injects a client-specific fingerprint vector $s_i$ into its local update $g_i$ before transmission:
\begin{equation}
\tilde{g}_i = g_i + \alpha_{\text{fp}} s_i,
\end{equation}
where $\alpha_{\text{fp}} > 0$ controls the fingerprint strength and $s_i \in \mathbb{R}^d$ is a sparse random vector generated using a client-specific seed, with support on $k \ll d$ randomly selected coordinates, zero-mean 
entries, and $\|s_i\| = 1$. The induced perturbation from client $i$ is $\delta_i = \alpha_{\text{fp}} s_i$. Under weighted FedAvg, the aggregated perturbation is:
\begin{equation}
\delta_{\text{global}} =
\alpha_{\text{fp}}
\sum_{i \in \mathcal{S}_{\text{det}}}
w_i s_i,
\quad
w_i = \frac{|\mathcal{D}_i|}{|\mathcal{D}_{\text{total}}|}.
\end{equation}
Expanding the squared norm yields:
\begin{equation}
\|\delta_{\text{global}}\|^2
=
\alpha_{\text{fp}}^2
\left(
\sum_{i \in \mathcal{S}_{\text{det}}} w_i^2 \|s_i\|^2
+
\sum_{i \neq j} w_i w_j \langle s_i, s_j \rangle
\right).
\end{equation}
Since fingerprint vectors are generated with independent random support of size $k \ll d$, the probability of support overlap is $\mathcal{O}(k^2/d)$. In this regime, cross inner products vanish in expectation, i.e., $\mathbb{E}[\langle s_i, s_j \rangle] \approx 0$ for $i \neq j$. Taking expectation eliminates the cross terms, yielding:
\begin{equation}
\mathbb{E}\|\delta_{\text{global}}\|^2
\approx
\alpha_{\text{fp}}^2
\sum_{i \in \mathcal{S}_{\text{det}}}
w_i^2.
\end{equation}
Since $\delta_{\text{global}}$ is a sum of independent sub-Gaussian or bounded-norm random vectors, its norm concentrates around its root-mean-square magnitude with high probability:
\begin{equation}
\|\delta_{\text{global}}\|
=
\mathcal{O}\left(
\alpha_{\text{fp}}
\sqrt{
\sum_{i \in \mathcal{S}_{\text{det}}}
w_i^2
}
\right).
\end{equation}
Under approximately equal dataset sizes ($w_i \approx 1/N$), letting $f_{\text{det}}^{(c)} = |\mathcal{S}_{\text{det}}|$ and $f_{\text{det}} = f_{\text{det}}^{(c)}/N$, we obtain $\sum_{i} w_i^2 = f_{\text{det}}/N$, and thus:
\begin{equation}
\|\delta_{\text{global}}\|
=
\mathcal{O}\left(
\alpha_{\text{fp}}
\sqrt{\frac{f_{\text{det}}}{N}}
\right).
\end{equation}

\paragraph{Effect on utility.} Assuming the global loss $F$ is $L$-smooth, the perturbation induces a change in the optimisation objective bounded by:
\begin{equation}
\Delta F
\leq
\frac{L}{2}\|\delta_{\text{global}}\|^2
=
\mathcal{O}
\left(
\alpha_{\text{fp}}^2
\frac{f_{\text{det}}}{N}
\right).
\end{equation}
This bounds the change in the optimisation objective rather than classification accuracy directly, as the 0--1 loss is non-smooth, and should therefore be interpreted as a proxy for utility degradation.

\paragraph{Client drift under FedAvg.} Each client performs $E \in [4,7]$ local epochs before aggregation, inducing client drift under heterogeneous data distributions. Following standard analyses of FedAvg, this 
introduces an additional deviation term $\epsilon_{\text{drift}} = \mathcal{O}(\eta^2 E^2 \Delta_{\text{grad}})$. The total deviation is therefore:
\begin{equation}
\Delta F
\leq
\mathcal{O}
\left(
\alpha_{\text{fp}}^2
\frac{f_{\text{det}}}{N}
\right)
+
\mathcal{O}(\eta^2 E^2 \Delta_{\text{grad}}).
\end{equation}

\paragraph{Regime interpretation.} In near-IID settings, $\Delta_{\text{grad}}$ is small and utility degradation is dominated by the fingerprint perturbation term, which scales as $\mathcal{O}(\alpha_{\text{fp}}^2 f_{\text{det}}/N)$ and decreases with increasing client 
count due to aggregation-induced cancellation. Under highly heterogeneous data distributions, the drift term dominates and masks the fingerprint perturbation effect. Under example settings ($\alpha_{\text{fp}} = 0.01$, 
$f_{\text{det}}^{(c)} = 5$, $N = 20$, $\eta = 0.005$, $E = 4$):
\begin{equation}
\|\delta_{\text{global}}\|
=
\mathcal{O}\left(
0.01 \cdot \frac{\sqrt{5}}{20}
\right)
\approx
\mathcal{O}(1.1 \times 10^{-3}),
\end{equation}
and
\begin{equation}
\Delta F
=
\mathcal{O}(1.2 \times 10^{-6})
+
\mathcal{O}(4 \times 10^{-4} \cdot \Delta_{\text{grad}}),
\end{equation}
suggesting that the perturbation-induced contribution to the optimisation objective remains small under the considered parameter regime.
\section{Utility analysis — additional datasets}
\label{sec:utility_additional}
\begin{figure*}[t]
    \centering
    \includegraphics[width=0.85\textwidth,height=8cm]{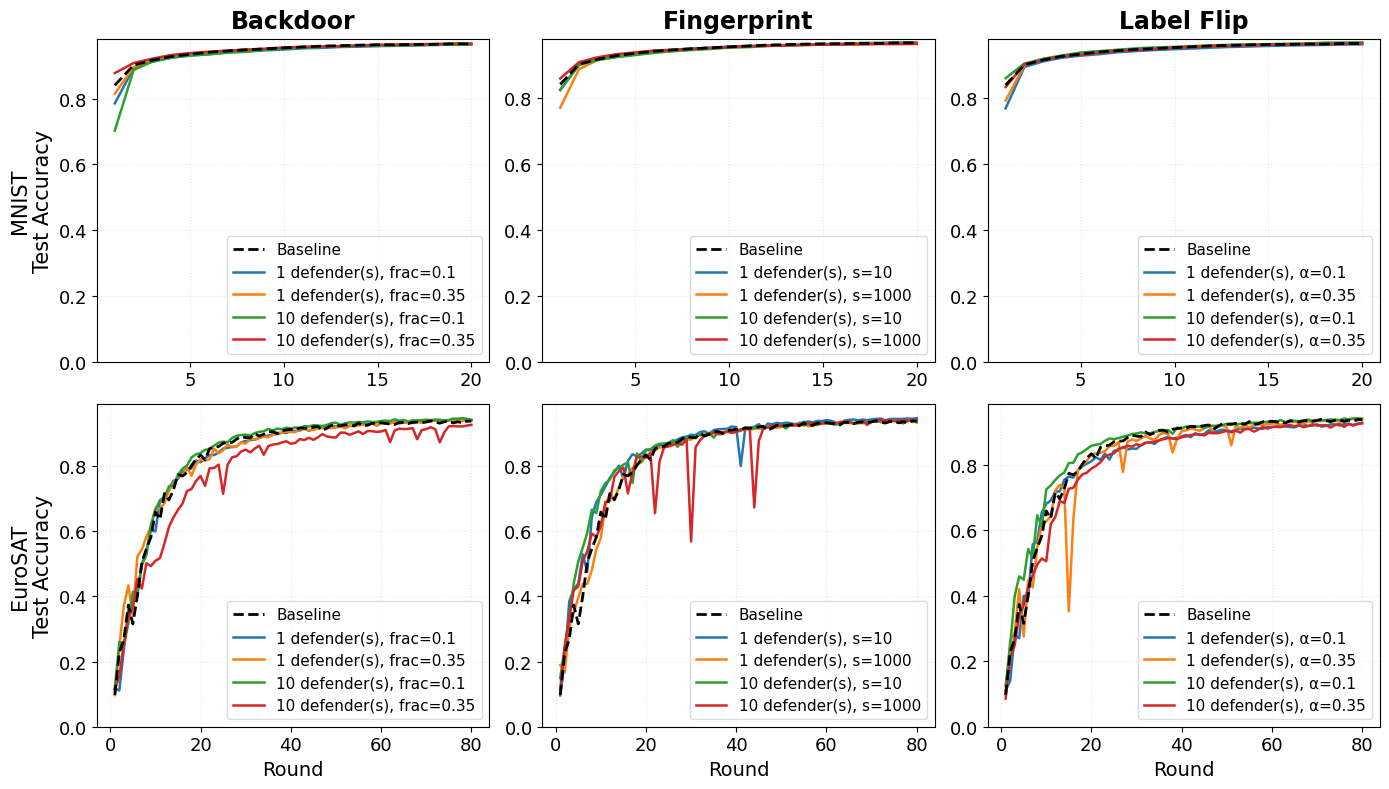}
    \caption{Orchestrator test accuracy across rounds for each detection method on MNIST (top row) and EUROSAT (bottom row), comparing the baseline against one and ten detecting clients under two parameter settings per method.}
    \label{fig:utility_mnist_eurosat}
\end{figure*}
We report utility results on MNIST and EUROSAT under the same experimental setup as Section~\ref{sec:utility} (20 clients, FedAvg, baseline vs one and ten detecting clients under two parameter settings per method). Figure~\ref{fig:utility_mnist_eurosat} reports orchestrator test accuracy across rounds for both datasets.
On MNIST (top row), all three methods produce negligible utility degradation -accuracy curves remain indistinguishable from the baseline across all parameter settings and detecting client counts. 

EUROSAT exhibits larger transient deviations, particularly for fingerprinting with ten detectors at $\alpha_{\text{fp}} = 9.1 \times 10^{-3}$ and for label flipping at $\alpha_{\text{flip}} = 0.35$. Backdoor trigger detection remains close to the baseline, with only a small persistent gap under the more aggressive setting with ten detectors. However, all methods converge near the baseline by later training rounds, indicating that the observed utility cost is driven primarily by amplified client drift and the harder optimisation dynamics of EUROSAT rather than persistent degradation from the detection mechanisms themselves.

These observations are consistent with the theoretical bounds in Appendix~\ref{sec:utility_bounds}.

\section{Performance analysis of label-flip and backdoor trigger methods at varying detection thresholds}
\label{Appendix_Best_Performing}
In this Section, we report the best-achieved accuracy and F1 Score for Label Flip (Algorithm \ref{alg:label_flipping}) and Backdoor Trigger (Algorithm \ref{alg:backdoor_trigger}) across all datasets and two testing scenarios (targeted client and targeted subset) together with a corresponding threshold. The results are combined within Table \ref{tab:best_thresholds_combined}. The corresponding visualisation of the results was placed in Figure \ref{fig:best_performance_comparison}.

The results directly correspond to the findings included in the main body of the paper. When a particular client is targeted (Scenario I), both methods of detection exhibit high Accuracy and F1 Score, irrespective of the threshold, with not a single error as reported on some datasets (Accuracy of 1.0 for MNIST and CIFAR10 datasets for the label flip detection method). This corresponds to a high reliability of the method for detecting the most dangerous (but also, simplest to detect) type of attack - one where there is no aggregation performed at all.

The attack is further complicated by the introduction of Scenario II, which targets a subset of clients. As demonstrated by Table \ref{tab:best_thresholds_combined}, Label Flip detection exhibits far better results than the Backdoor Trigger, with accuracy ranging from $0.6$ to $0.93$ depending on the dataset and F1 Score in the range of $0.72$ to $0.93$. Although not directly explored in this work, the worst performance of the Backdoor Trigger may be directly associated with how the trigger is embedded in the training data. Future research could explore how different trigger placement methods may impact the method's performance.

%\begin{table*}[ht!]
  %\centering
  %\footnotesize
  %\renewcommand{\arraystretch}{1.15}
  %\begin{tabular}{llcccccccc}

\begin{table*}[ht!]
  \centering
  \footnotesize
  \renewcommand{\arraystretch}{1.15}
  %\resizebox{\textwidth}{!}{%
   \scalebox{1.2}{%
  
  \begin{tabular}{llcccccccc}
    \toprule
    \multirow{2}{*}{Detection Type} &
    \multirow{2}{*}{Dataset} & 
    \multicolumn{2}{c}{Scenario I Accuracy} & 
    \multicolumn{2}{c}{Scenario I F1 Score} & 
    \multicolumn{2}{c}{Scenario II Accuracy} & 
    \multicolumn{2}{c}{Scenario II F1 Score} \\
    \cmidrule(lr){3-4} \cmidrule(lr){5-6} \cmidrule(lr){7-8} \cmidrule(lr){9-10}
    & & Value & Threshold & Value & Threshold & Value & Threshold & Value & Threshold \\
    \midrule
    \multirow{5}{*}{Label Flip} 
    & MNIST     & 1.000 & $< 0.01$ & 1.000 & $< 0.01$ & 0.611 & $< 0.3$  & 0.720 & $< 0.95$ \\
    & CIFAR10   & 1.000 & $< 0.01$ & 1.000 & $< 0.01$ & 0.938 & $< 0.1$  & 0.939 & $< 0.1$  \\
    & CIFAR100  & 0.931 & $< 0.01$ & 0.935 & $< 0.01$ & 0.655 & $< 0.3$  & 0.737 & $< 0.3$  \\
    & PATHMNIST & 0.970 & $< 0.01$ & 0.970 & $< 0.01$ & 0.697 & $< 0.9$  & 0.762 & $< 0.9$  \\
    & EUROSAT   & 0.974 & $< 0.01$ & 0.974 & $< 0.01$ & 0.711 & $< 0.5$  & 0.776 & $< 0.5$  \\
    \midrule
    \multirow{5}{*}{Backdoor Trigger}
    & MNIST     & 0.950 & $> 0.99$ & 0.947 & $> 0.99$ & 0.550 & $> 0.5$  & 0.667 & $> 0.01$ \\
    & CIFAR10   & 0.980 & $> 0.7$  & 0.980 & $> 0.7$  & 0.520 & $> 0.3$  & 0.658 & $> 0.01$ \\
    & CIFAR100  & 0.983 & $> 0.9$  & 0.983 & $> 0.9$  & 0.583 & $> 0.7$  & 0.659 & $> 0.01$ \\
    & PATHMNIST & 0.971 & $> 0.99$ & 0.970 & $> 0.99$ & 0.543 & $> 0.99$ & 0.625 & $> 0.01$ \\
    & EUROSAT   & 0.975 & $> 0.95$ & 0.974 & $> 0.95$ & 0.600 & $> 0.1$  & 0.704 & $> 0.1$  \\
    \bottomrule
  \end{tabular}}
  \caption{Best performing accuracy and F1 Score thresholds $\alpha$ by dataset across scenarios for Label Flip (Algorithm \ref{alg:label_flipping}) and Backdoor Trigger (Algorithm \ref{alg:backdoor_trigger}) attacks. For each reported value (Accuracy or F1 Score) achieved on a dataset in a given scenario, a corresponding detection threshold is reported. The corresponding data is plotted on Figures \ref{fig:label_flip_best} and \ref{fig:backdoor_best}.} 
  \label{tab:best_thresholds_combined}
\end{table*}

\FloatBarrier
\section{Threshold impact on label flip and backdoor trigger algorithms.}
\label{Appendix_Threshold_Impact}
\begin{figure*}
    \centering
    \begin{subfigure}[t]{\linewidth}
        \centering
        \includegraphics[width=0.8\linewidth,height=9cm]{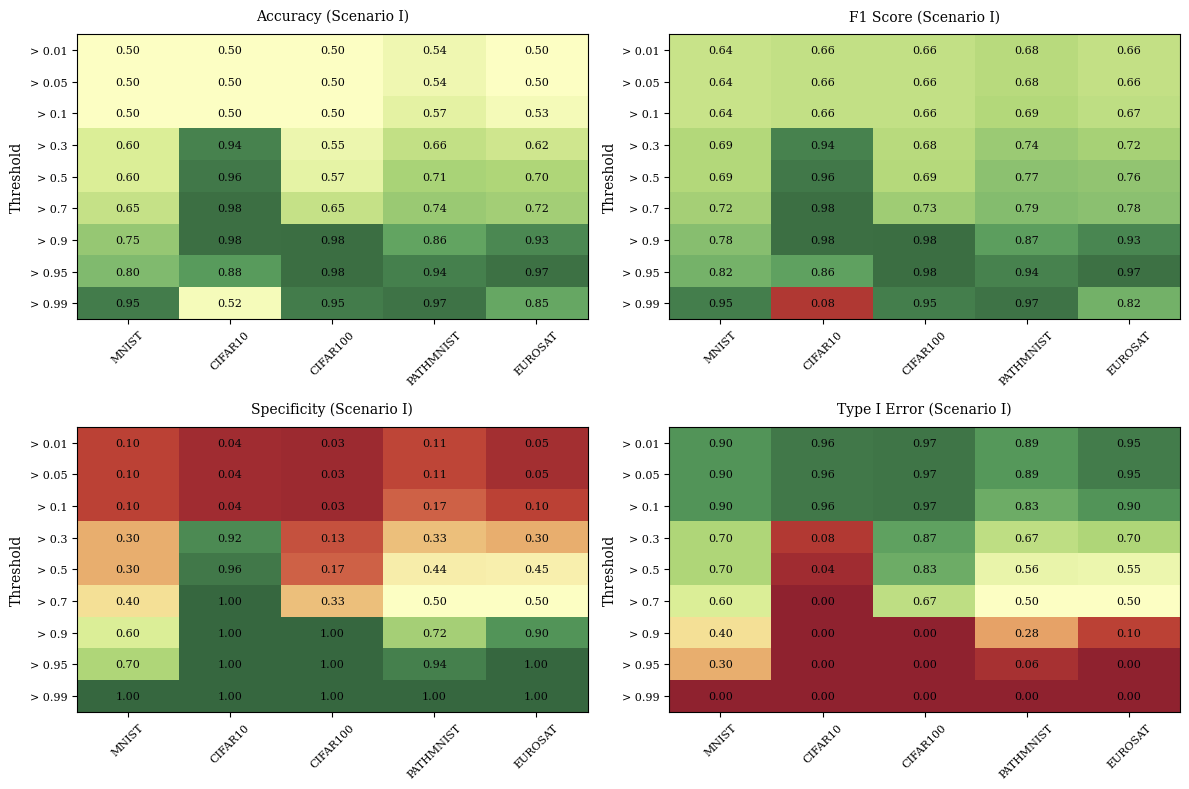}
        \caption{Metric variability  across thresholds $\theta$ for the backdoor trigger detection (Algorithm \ref{alg:backdoor_trigger}) for Scenario I}
       
        \label{fig:backdoor_threshold_s1}
    \end{subfigure}
    
    \vspace{0.5em}
    
    \begin{subfigure}[t]{\linewidth}
        \centering
        \includegraphics[width=0.8\linewidth,height=9cm]{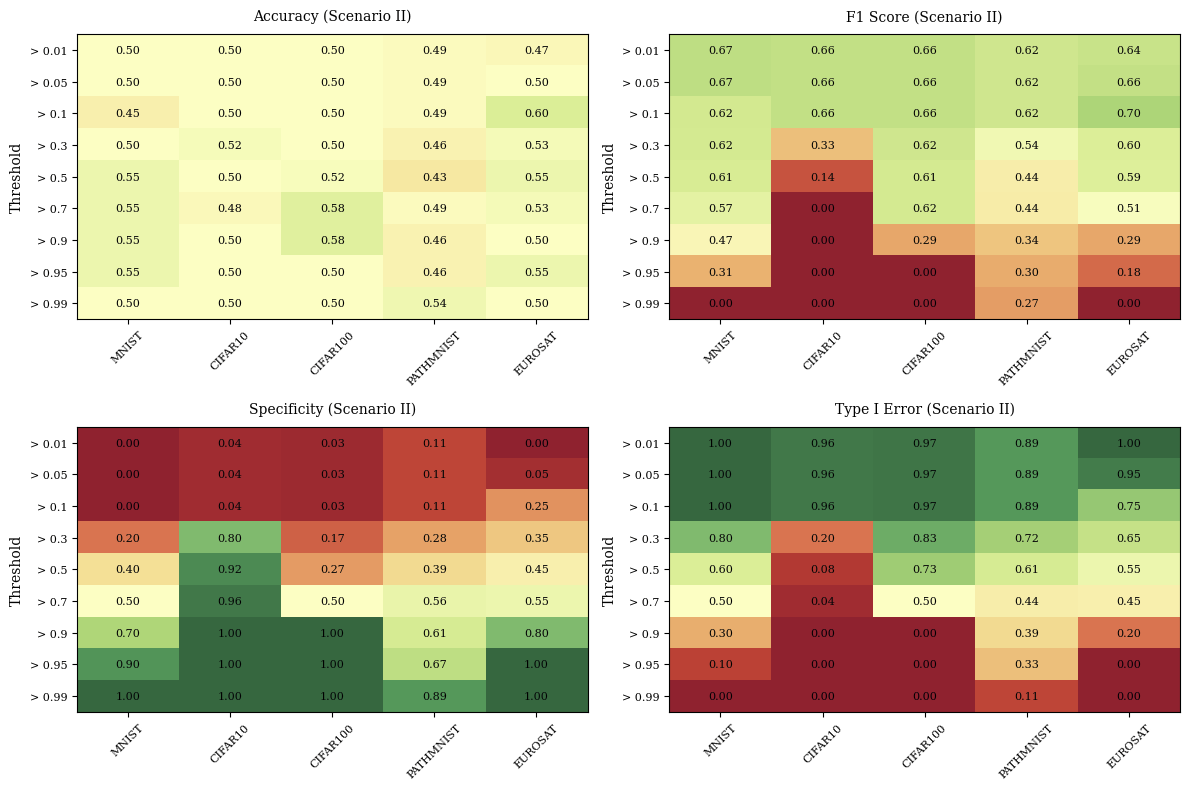}
        \caption{Metric variability  across thresholds $\theta$ for the backdoor trigger detection (Algorithm \ref{alg:backdoor_trigger}) For Scenario II}
      
        \label{fig:backdoor_threshold_s2}
    \end{subfigure}
    
    \caption{Backdoor Trigger (Algorithm \ref{alg:backdoor_trigger}) attack detection performance: Detection performance metrics across various detection thresholds $\theta$ showing how threshold selection affects accuracy, precision, recall, and F1-score across different datasets. Sub-figure \ref{fig:backdoor_threshold_s1} overviews Scenario I (overfit of a particular client), while the sub-figure \ref{fig:backdoor_threshold_s2} overviews Scenario II (overfit of a subset of clients).}
    \label{fig:backdoor_threshold_analysis}
\end{figure*}

\begin{figure*}
    \centering
    \begin{subfigure}[t]{\linewidth}
        \centering
        \includegraphics[width=0.8\linewidth,height=9cm]{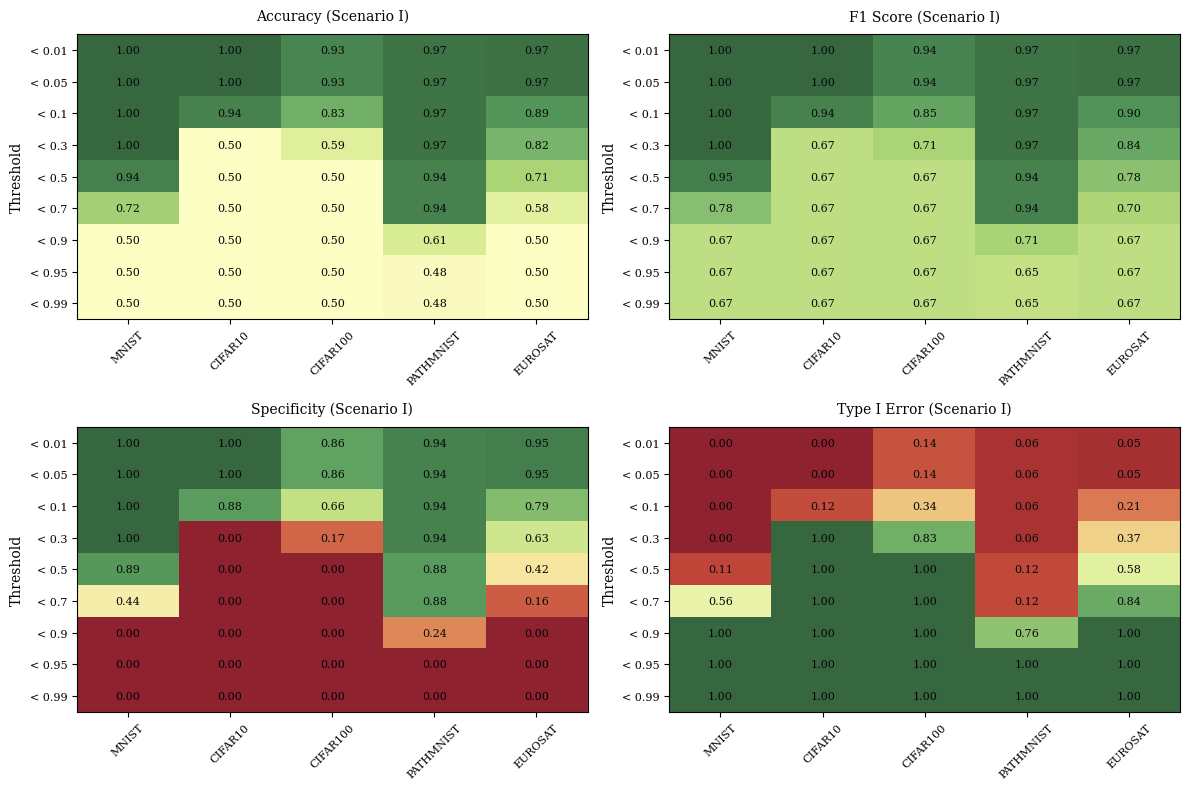}
        \caption{Scenario I:  Metric variability  across thresholds $\tau_{pes}$ for the label flip detection (Algorithm \ref{alg:label_flipping}}
      
        \label{fig:labelflip_threshold_s1}
    \end{subfigure}
    
    \vspace{1em}
    
    \begin{subfigure}[t]{\linewidth}
        \centering
        \includegraphics[width=0.8\linewidth,height=9cm]{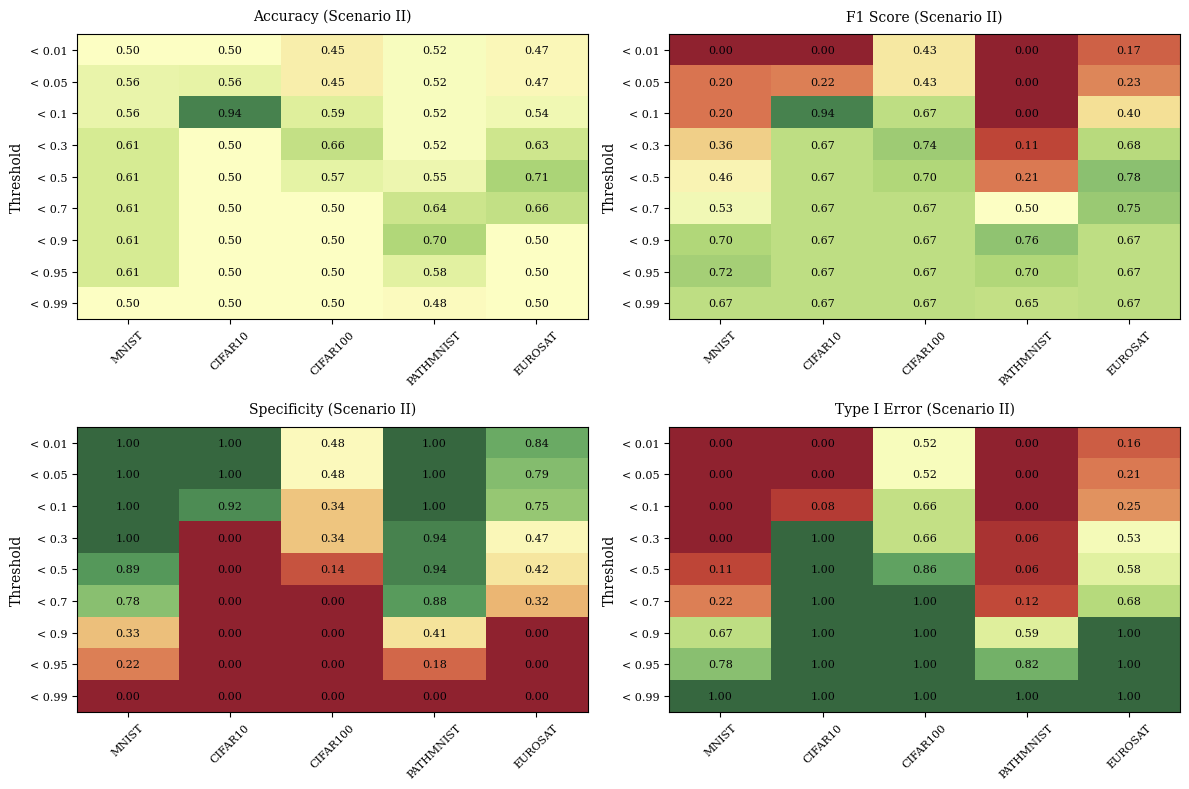}
        \caption{Scenario II:  Metric variability  across thresholds $\alpha$ for the label flip detection (Algorithm \ref{alg:label_flipping}}
       
        \label{fig:labelflip_threshold_s2}
    \end{subfigure}
    
    \caption{Label Flip (Algorithm \ref{alg:label_flipping}) attack detection performance: Detection performance metrics across various detection thresholds $\tau_{pes}$ showing how threshold selection affects accuracy, precision, recall, and F1-score across different datasets. Sub-figure \ref{fig:labelflip_threshold_s1} overviews Scenario I (overfit of a particular client), while the sub-figure \ref{fig:labelflip_threshold_s2} overviews Scenario II (overfit of a subset of clients).}
    \label{fig:label_threshold_analysis}
\end{figure*}

This Section examines how different detection thresholds affect the performance of the Label Flip (Algorithm \ref{alg:label_flipping}) and the Backdoor Trigger (Algorithm \ref{alg:backdoor_trigger}). Performance is measured and reported as accuracy, F1 Score, Specificity, and Type I Error across all five datasets and two attack scenarios. Figure \ref{fig:backdoor_threshold_analysis} presents the results for Backdoor Trigger, while Figure \ref{fig:label_threshold_analysis} presents the results for the Label Flip detection method. 

The results are consistent with previously reported data on the performance of both algorithms, indicating higher performance for individual targeting (Scenario I) and lower performance for a subset targeting (Scenario II), with the Backdoor Trigger detection method outperforming the Label Flip when the actual subset of clients is targeted.

\section{Additional results of the non-IID impact on the verification performance}
\label{appendix:non_iid_additional}

\begin{figure}
    \centering
    \includegraphics[width=1\linewidth]{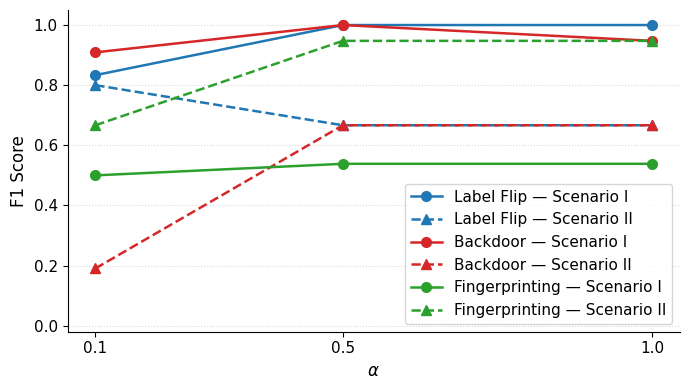}
    \caption{Detection F1Score (first 10 rounds) varying by the parameter $\alpha$ for all three methods on the EUROSAT dataset. F1Score is reported for only the best-performing threshold for a given $\alpha$ value. Thresholds are reported in Table \ref{tab:noniid-thresholds}.}
    \label{fig:eurosat-alpha}
\end{figure}

In this Section, we report additional experiments performed on the EUROSAT dataset (Figure \ref{fig:eurosat-alpha}) and the table of thresholds $\tau$ (Table \ref{tab:noniid-thresholds}) that were used for calculating the F1Score of detection methods as presented in Figure \ref{fig:eurosat-alpha} and \ref{fig:mnist-alpha}.

\begin{table*}[ht]
\centering

\label{tab:hyperparams}
\begin{tabular}{|l|l|l|c|c|c|}
\toprule
Dataset & Method & Scenario & $\alpha{=}0.1$ & $\alpha{=}0.5$ & $\alpha{=}1.0$ \\
\midrule
EUROSAT & Backdoor & Scenario I & 0.05 & 0.10 & 0.10 \\
EUROSAT & Backdoor & Scenario II & 0.01 & 0.30 & 0.05 \\
EUROSAT & Fingerprinting & Scenario I & -1.00 & -1.00 & -1.00 \\
EUROSAT & Fingerprinting & Scenario II & -1.00 & -1.00 & -1.00 \\
EUROSAT & Label-Flip & Scenario I & 0.01 & 0.01 & 0.01 \\
EUROSAT & Label-Flip & Scenario II & 0.50 & 0.05 & 0.50 \\
MNIST & Backdoor & Scenario I & 0.05 & 0.10 & 0.30 \\
MNIST & Backdoor & Scenario II & 0.01 & 0.01 & 0.01 \\
MNIST & Fingerprinting & Scenario I & -1.00 & 0.50 & 0.50 \\
MNIST & Fingerprinting & Scenario II & -1.00 & 0.50 & 0.50 \\
MNIST & Label-Flip & Scenario I & 0.01 & 0.01 & 0.01 \\
MNIST & Label-Flip & Scenario II & 0.30 & 0.50 & 0.70 \\
\bottomrule
\end{tabular}
\caption{Threshold $\tau$ used for particular runs (depending on the $\alpha$ value) as displayed in Figures \ref{fig:mnist-alpha} or \ref{fig:eurosat-alpha}.}
\label{tab:noniid-thresholds}
\end{table*}

\section{Additional results for scalability analysis}
\label{appendix:scalability_expanded}
\begin{figure*}
    \centering
    \includegraphics[width=1\linewidth,height=10cm]{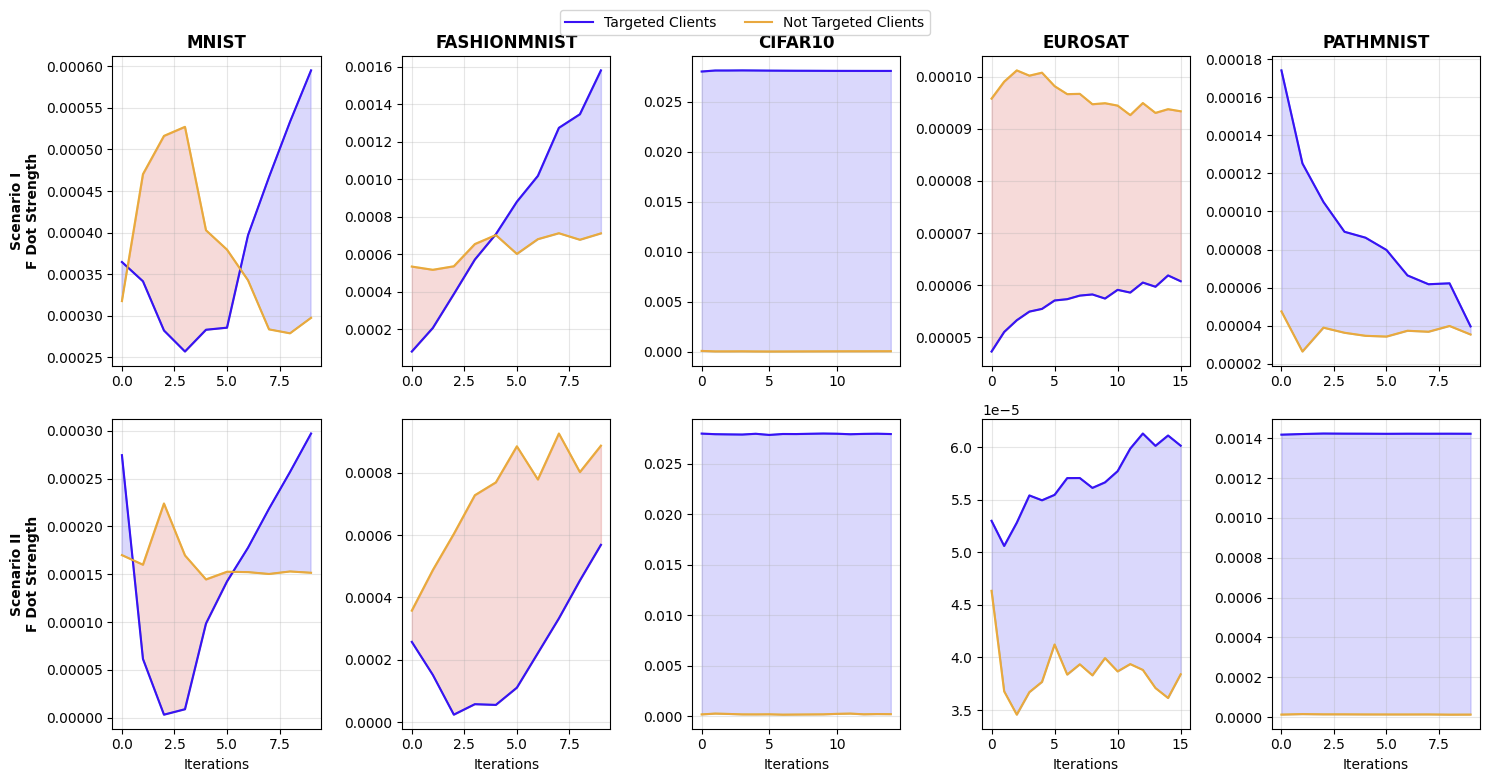}
    \caption{Fingerprint strength for FedOpt (20 clients) across all five datasets. The dot product value between a dispatched (local) and received (aggregated) model is plotted on the y-axis, while the x-axis represents the iteration of the global training. The columns show Scenario I (first) and Scenario II (second), while the rows represent the datasets. The coloured line shows the dot strength of the signed and received models per epoch, averaged across multiple verified clients when present. The area between targeted and non-targeted clients is blue if the fingerprint of the targeted model is stronger than its non-targeted counterpart (expected) and red otherwise.}
    \label{fig:fingerprint_20_fedopt}
\end{figure*}

\begin{figure*}[ht]
    \centering
    \includegraphics[width=0.85\linewidth,height=14cm]{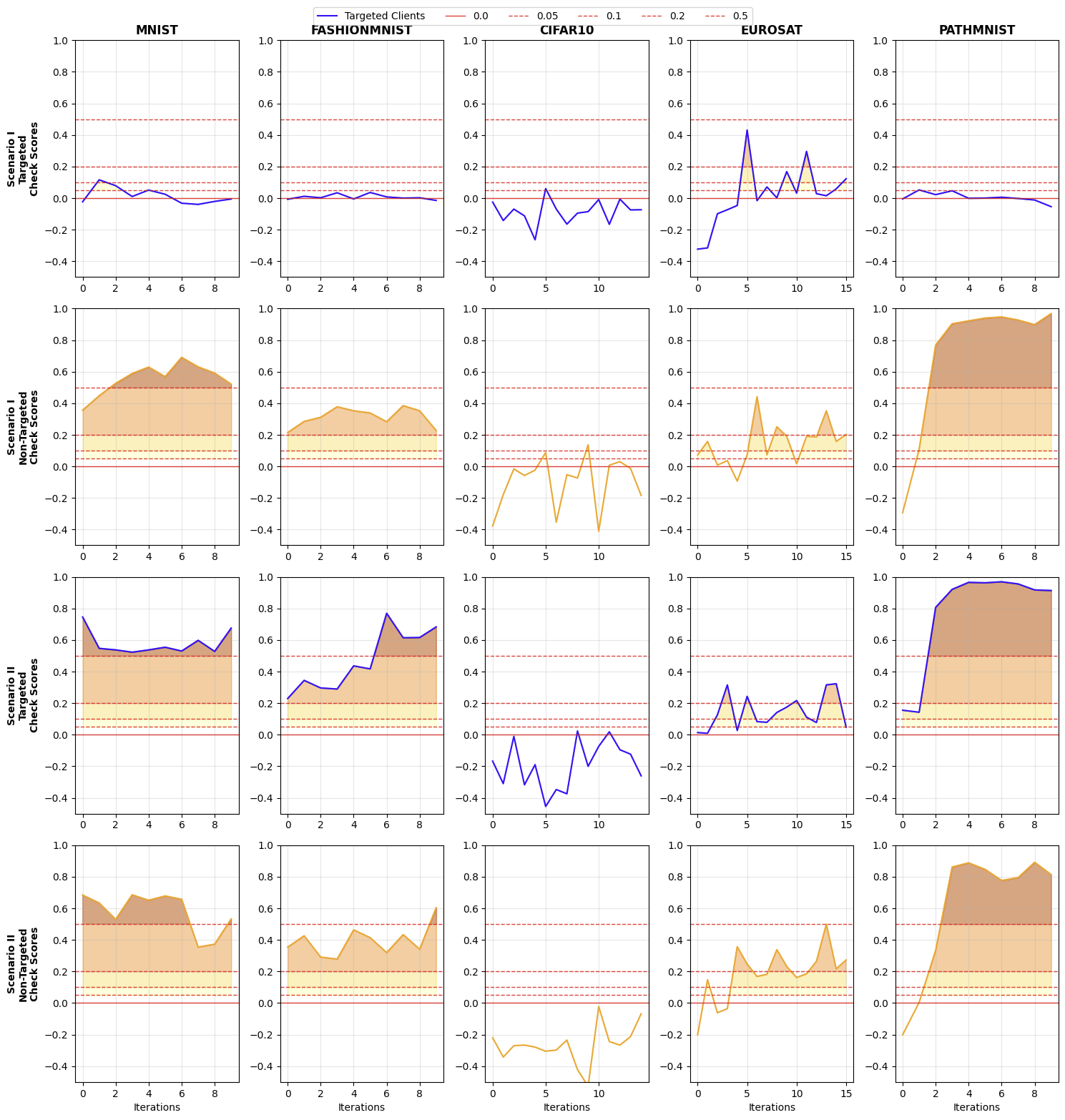}
    \caption{PES score for Label Flip (FedOpt, 20 clients) across all five datasets. The line shows the PES value (Algorithm \ref{alg:label_flipping}) per epoch; if multiple verified clients exist, the line represents their average. The PES value is plotted on the y-axis, while the iteration is placed on the x-axis. Dotted red lines indicate thresholds (0.0, 0.05, 0.1, 0.2, 0.5), and the colored region highlights scores above the threshold. Rows represent particular scenarios and targeted and non-targeted subsets (the first two rows overview scores registered for targeted and non-targeted clients for Scenario I, while the third and fourth rows represent the scores registered for targeted clients for Scenario II). Columns represent particular datasets. By design, non-targeted clients (second and fourth rows) are expected to exceed the thresholds, while targeted clients (first and third rows) are expected to remain below the thresholds.
    }
    \label{fig:label_flip_20_fedopt}
\end{figure*}

\begin{figure*}
    \centering
    \includegraphics[width=0.9\linewidth,height=16cm]{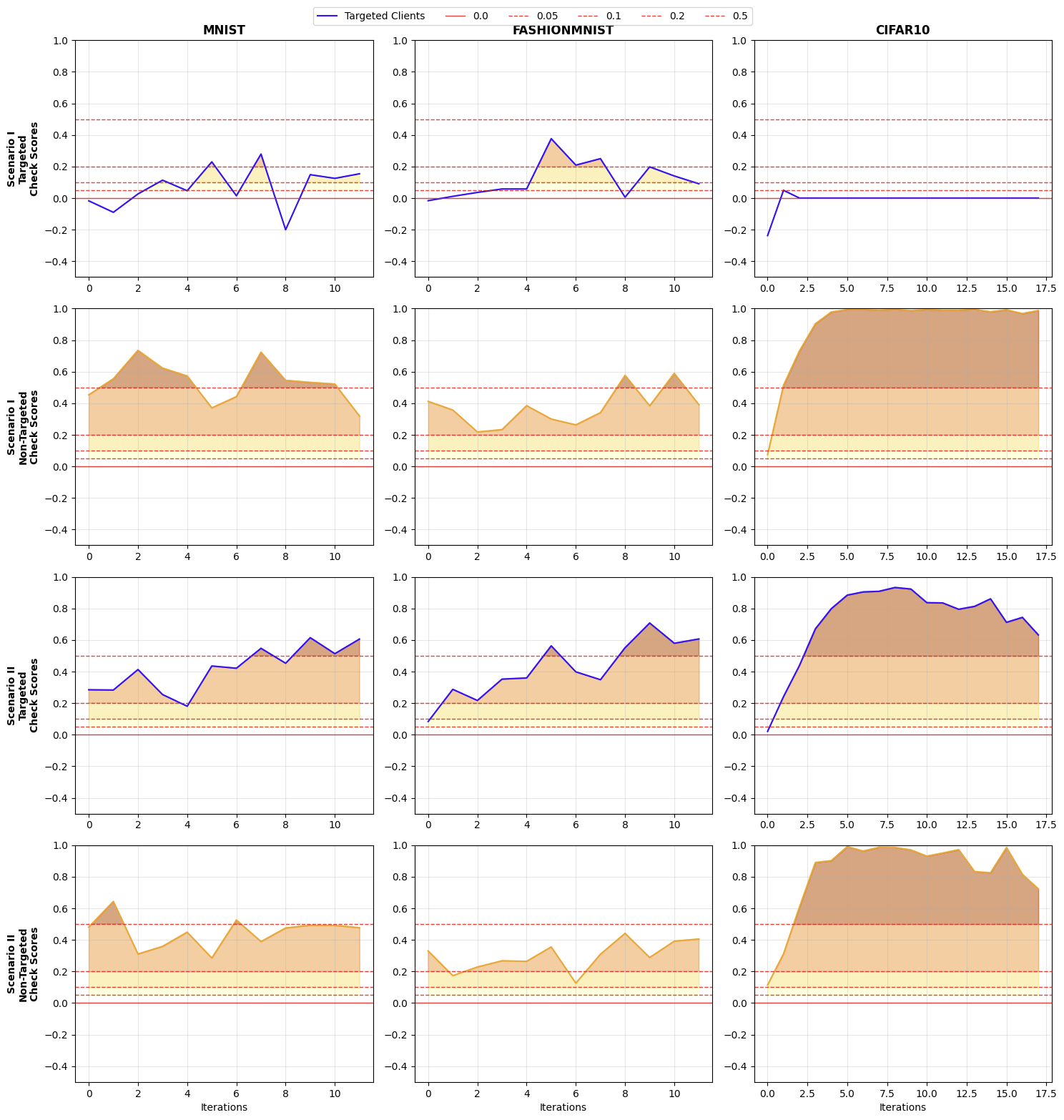}
    \caption{PES score for Label Flip (FedOpt, 50 clients) across all five datasets. The line shows the PES value (Algorithm \ref{alg:label_flipping}) per epoch; if multiple verified clients exist, the line represents their average. The PES value is plotted on the y-axis, while the iteration is placed on the x-axis. Dotted red lines indicate thresholds (0.0, 0.05, 0.1, 0.2, 0.5), and the colored region highlights scores above the threshold. Rows represent particular scenarios and targeted and non-targeted subsets (the first two rows overview scores registered for targeted and non-targeted clients for Scenario I, while the third and fourth rows represent the scores registered for targeted clients for Scenario II). Columns represent particular datasets. By design, non-targeted clients (second and fourth rows) are expected to exceed the thresholds, while targeted clients (first and third rows) are expected to remain below the thresholds.
    }
    \label{fig:label_flip_50_fedopt}
\end{figure*}

\begin{figure*}
    \centering
    \includegraphics[width=1\linewidth,height=7cm]{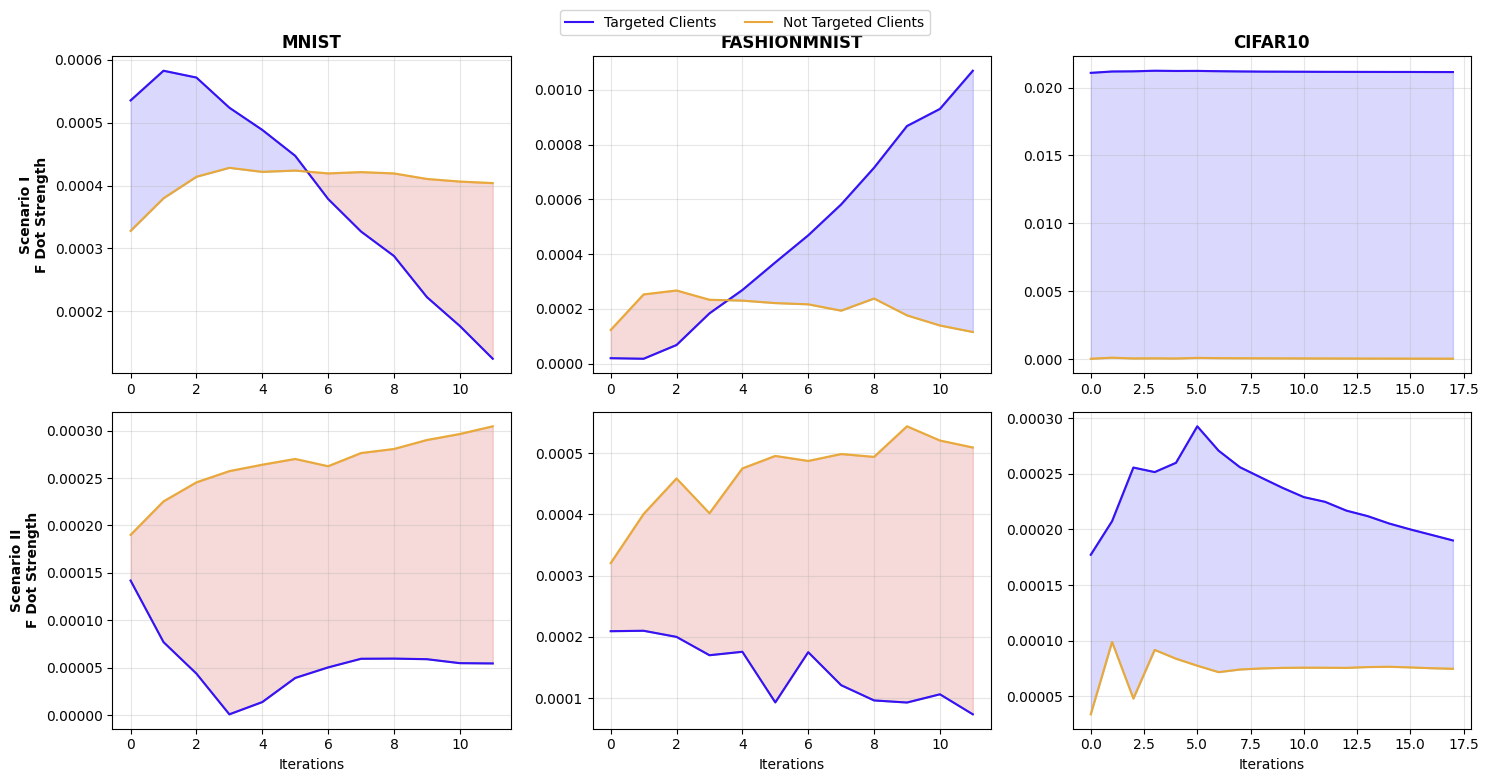}
    \caption{Fingerprint strength for FedOpt (50 clients) across all five datasets. The dot product value between a dispatched (local) and received (aggregated) model is plotted on the y-axis, while the x-axis represents the iteration of the global training. The columns show Scenario I (first) and Scenario II (second), while the rows represent the datasets. The coloured line shows the dot strength of the signed and received models per epoch, averaged across multiple verified clients when present. The area between targeted and non-targeted clients is blue if the fingerprint of the targeted model is stronger than its non-targeted counterpart (expected) and red otherwise.}
    \label{fig:fingerprint_50_fedopt}
\end{figure*}

\begin{figure*}
    \includegraphics[width=1\linewidth,height=8cm]{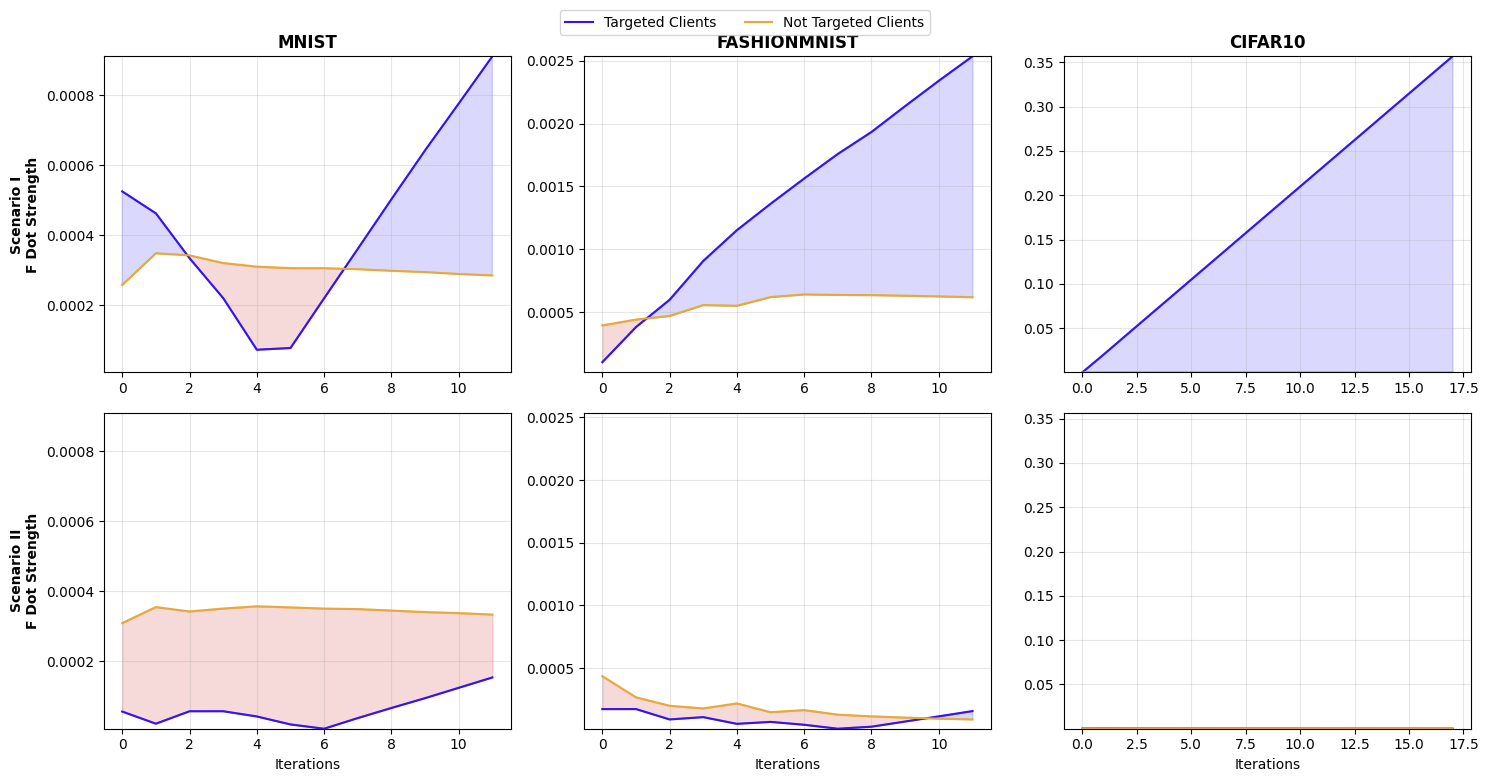}
    \caption{Fingerprint strength for FedAvg (50 clients) across all five datasets. The dot product value between a dispatched (local) and received (aggregated) model is plotted on the y-axis, while the x-axis represents the iteration of the global training. The columns show Scenario I (first) and Scenario II (second), while the rows represent the datasets. The coloured line shows the dot strength of the signed and received models per epoch, averaged across multiple verified clients when present. The area between targeted and non-targeted clients is blue if the fingerprint of the targeted model is stronger than its non-targeted counterpart (expected) and red otherwise.}
    \label{fig:fingerprint_50_fedavg}
\end{figure*}

\begin{figure*}
    \centering
    \includegraphics[width=0.9\linewidth,height=16cm]{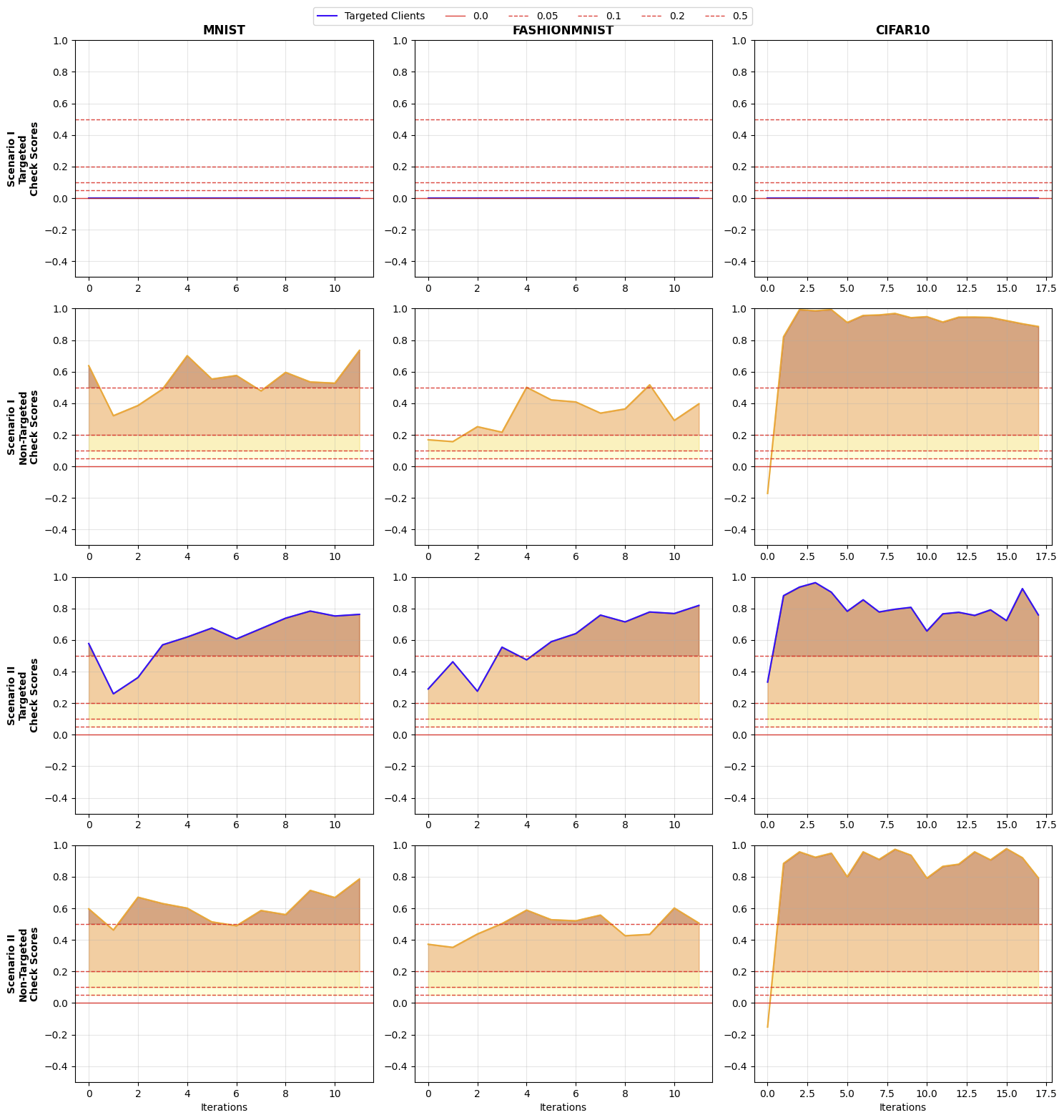}
    \caption{PES score for Label Flip (FedAvg, 50 clients) across all five datasets. The line shows the PES value (Algorithm \ref{alg:label_flipping}) per epoch; if multiple verified clients exist, the line represents their average. The PES value is plotted on the y-axis, while the iteration is placed on the x-axis. Dotted red lines indicate thresholds (0.0, 0.05, 0.1, 0.2, 0.5), and the colored region highlights scores above the threshold. Rows represent particular scenarios and targeted and non-targeted subsets (the first two rows overview scores registered for targeted and non-targeted clients for Scenario I, while the third and fourth rows represent the scores registered for targeted clients for Scenario II). Columns represent particular datasets. By design, non-targeted clients (second and fourth rows) are expected to exceed the thresholds, while targeted clients (first and third rows) are expected to remain below the thresholds.
    }
    \label{fig:label_flip_50_fedavg}
\end{figure*}

This Section examines the scalability of the presented solutions by reporting results for both $20$ and $50$ clients across two attack scenarios and two aggregation protocols (FedAvg and FedOpt). Figures \ref{fig:fingerprint_20_fedavg} and \ref{fig:fingerprint_20_fedopt} present the experiments on fingerprint strength for FedAvg and FedOpt aggregation methods, respectively (20 clients). The replicated experiments for 50 clients are shown in Figure \ref{fig:fingerprint_50_fedavg} and \ref{fig:fingerprint_50_fedopt}. Figures \ref{fig:label_flip_20_fedavg} and \ref{fig:label_flip_20_fedopt} display the PES score for Label Flip for FedAvg and FedOpt aggregation methods, respectively (20 clients). The replicated experiments for 50 clients are shown in Figure \ref{fig:label_flip_50_fedavg} and \ref{fig:label_flip_50_fedopt}.

\FloatBarrier
\FloatBarrier

\end{document}

%% file: macro.tex
\usepackage{xspace}
\usepackage{soul}
\usepackage{etoolbox}
\newtoggle{trackchanges}
\toggletrue{trackchanges}
\usepackage{adjustbox}
\usepackage{supertabular,booktabs}

\newcommand{\etal}{\textit{et al.}\xspace}
\newcommand{\eg}{\textit{e.g.},\xspace}
\newcommand{\ie}{\textit{i.e.},\xspace}